\documentclass[conference]{IEEEtran}
\IEEEoverridecommandlockouts
\usepackage{cite}
\usepackage{amsmath,amssymb,amsfonts}
\usepackage{graphicx}
\usepackage{textcomp}
\usepackage{xcolor}
\usepackage{multirow}
\usepackage{amsmath}
\usepackage{algorithm}
\usepackage[noend]{algpseudocode}
\usepackage{amsfonts}

\begin{document}

\title{Network Attack Traffic Detection With Hybrid Quantum-Enhanced Convolution Neural Network\\}


\author{\IEEEauthorblockN{Zihao Wang, Kar-Wai Fok, Vrizlynn L. L. Thing}
\IEEEauthorblockA{\textit{Cybersecurity Strategic Technology Centre} \\
\textit{ST Engineering}\\
Singapore, Singapore \\
\{zihao.wang, fok.karwai\}@stengg.com, vriz@ieee.org}
}
\maketitle                       

\section{\bf Abstract}
The emerging paradigm of Quantum Machine Learning (QML) combines features of quantum computing and machine learning (ML). QML enables the generation and recognition of statistical data patterns that classical computers and classical ML methods struggle to effectively execute. QML utilizes quantum systems to enhance algorithmic computation speed and real-time data processing capabilities, making it one of the most promising tools in the field of ML. Quantum superposition and entanglement features also hold the promise to potentially expand the potential feature representation capabilities of ML. Therefore, in this study, we explore how quantum computing affects ML and whether it can further improve the detection performance on network traffic detection, especially on unseen attacks which are types of malicious traffic that do not exist in the ML training dataset. Classical ML models often perform poorly in detecting these unseen attacks because they have not been trained on such traffic. Hence, this paper focuses on designing and proposing novel hybrid structures of Quantum Convolutional Neural Network (QCNN) to achieve the detection of malicious traffic. The detection performance, generalization, and robustness of the QML solutions are evaluated and compared with classical ML running on classical computers. The emphasis lies in assessing whether the QML-based malicious traffic detection outperforms classical solutions. Based on experiment results, QCNN models demonstrated superior performance compared to classical ML approaches on unseen attack detection.

\begin{IEEEkeywords}
quantum machine learning, encrypted malicious traffic detection, traffic classification, deep learning, traffic analysis, 5G network.
\end{IEEEkeywords}

\section{\bf Introduction}
In recent years, the widespread adoption of emerging technologies such as the Meta-verse, augmented reality (AR), virtual reality (VR), cloud computing, and the rapid and constant evolution of connectivity has massively increased the traffic demand. According to the new Nokia Global Network Traffic 2030 report~\cite{AA11}, there is an expected compound annual growth rate of 22\% to 25\% in the demand for global telecom networks from 2022 to 2033. Even in a moderate scenario, the numbers involved are still huge. Network traffic is expected to grow between 4 to 9 times by 2030.

In order to safeguard the rapidly growing network environment, traditional Intrusion Detection Systems (IDS) that rely on Deep Packet Inspection (DPI) techniques~\cite{AA12} are no longer effective, especially with the advent of traffic encryption mechanisms. As a result, machine learning methods have been successfully used to detect security intrusions and malicious activities as an alternative to traditional DPI detection methods~\cite{AA13}. However, with the explosive growth of network traffic data in recent years, machine learning technology has also encountered new bottlenecks when dealing with massive amounts of input traffic and increasing unseen malicious traffic.

Massive amounts of network traffic data, diverse types of malicious attacks, and the continuous emergence of unseen malicious traffic have forced researchers to continuously expand the scale of classical machine learning datasets. This leads to prolonged training and detection times for model algorithms. Existing detection models struggle to promptly identify malicious traffic and unseen traffic within large-scale network data. On the other hand, using small datasets for fast training and computation may lead to reduced model accuracy and insufficient generalization capabilities. Both approaches diverge from the characteristics of malicious activity, such as the variability, intensity, and adaptability of intrusions~\cite{AA14}. 

Quantum Machine Learning (QML) represents the fusion of quantum computing with machine learning solutions. Leveraging quantum resources like entanglement and state superposition, it holds promise as a potential solution to these challenges. QML utilizes quantum systems to enhance algorithmic computation speed and real-time data processing capabilities, addressing the detection of massive network data~\cite{AA15}~\cite{AA16}~\cite{AA17}. Moreover, by embedding network traffic data into a high-dimensional quantum feature space for learning, models are enabled to recognize patterns that classical methods might overlook. Therefore, QML may have the potential to improve the performance in identifying unseen malicious traffic.

However,  there is limited research applying QML in the field of malicious traffic detection. Existing QML in network intrusion detection also suffers from several limitations, including a lack of QML algorithm design, limited comparisons between QML and classical machine learning counterpart models, outdated datasets, and insufficient exploration of classical-to-quantum data conversion.

Combining quantum computing and machine learning features enables the generation and recognition of statistical data patterns that cannot be effectively performed by classical computers and classical machine learning. Thus, we proposed to design hybrid quantum machine learning (QML) models to study and evaluate the impact of quantum in improving classical methods. The focus of this paper is on designing a hybrid quantum convolutional neural network (Quantum CNN) structure to achieve malicious traffic detection. The contributions of this paper are as follows:

1. Design 5 different architectures of hybrid Quantum CNN models with different quantum embedding methods, quantum computing algorithms, and positions of quantum layers in hybrid models.

2. Conduct comparative experiments among different hybrid Quantum CNN models and classical CNN model.

3. Evaluate the detection capabilities of different hybrid Quantum CNN models on unseen data so as to test the generalization and robustness of the model.

The organization of the rest of this paper is as follows: Section II presents the literature review of different detection techniques for malicious traffic detection and analysis including quantum machine learning based traffic detection approaches. Then the methodology of the hybrid quantum CNN model is proposed in Section III.  In Section IV, we present the setup of our experiments and conduct the performance evaluations. We conclude the paper in Section V, by discussing the remaining challenges and future directions.

\section{\bf Literature Review}

\subsection{\bf Classical Malicious Traffic Detection}
Bader et al.~\cite{AA19} proposed MalDIST based on the extension of the DISTILLER model~\cite{AA24} to encrypted malicious traffic detection and classification. The proposed framework consists of several deep learning models, including 1D CNN and 2D CNN. 99.7\% accuracy, precision, recall, and F1 are achieved. A network traffic graph neural network model (NT-GNN) is proposed by Liu et al.~\cite{AA20} The model considers the node and edge aspects of the graph, capturing the connections between various traffic flows and individual traffic features. The model achieves 97\% accuracy in the CICAndMal2017 and AAGM datasets. Zheng et al.~\cite{AA21} pointed out that the statistical feature-based method focuses on the internal information of the network flows and the graph-based method focuses on the external connections between network flows. The authors considered both internal information and external connections to propose a graph convolutional network model (GCN-ETA). GCN-ETA consists of a modified GCN based feature extractor and a decision tree classifier to effectively improve the effectiveness and speed of encrypted malicious traffic detection. The experiment shows the method can perform 98\% higher accuracy, AUC, and F1 scores, and achieve more than 1300 traffic flows per second. Liu et al.~\cite{AA22} combines spatial-temporal features with a dual-attention mechanism. The proposed model contains four parts: spatial feature learning by using 1D-CNN and BiGRU, attention mechanism based on encrypted packet, flow temporal feature learning between encrypted streams, and attention mechanism based on flow. The approach provides rich encrypted traffic characteristics and insights for encrypted malicious traffic detection and classification.

Classical machine learning has achieved excellent results in malicious traffic detection. However, there are still two unsolved issues. First, most experiments in research rely on a limited number of publicly available offline training and testing datasets~\cite{AA20}~\cite{AA21}~\cite{AA22}~\cite{AA23}. When dealing with a large amount of continuous online traffic, both the training time and the detection time increase. The second issue is related to the poor performance on unseen attacks. Existing models may lack sufficient detection capabilities~\cite{AA13} when faced with emerging malicious activities. Many previous studies even did not conduct unseen malicious traffic detection evaluation. Therefore, the researchers introduced QML to overcome the challenges posed by big data and classical model detection to ensure high-performance malicious traffic detection while maintaining accuracy in detecting unseen malicious traffic.

\subsection{\bf Quantum Machine Learning (QML) based Network Traffic Detection}
Payares et al.~\cite{AA36} present three quantum models to detect distributed denial of service attacks. They compare Quantum Support Vector Machines, hybrid Quantum Classical Neural Networks, and ensemble above two models as an ensemble detection model. The comparative experiment for these three models is conducted with CIC-DDoS2019 dataset. The experiment result shows that detecting DDoS type threats is possible using QML with high accuracy. Kadry et al~\cite{AA28} proposed a QNN with Wale Optimization algorithm(WOA) intrusion detection system framework. The authors selected KDD-CUP 99 dataset and achieved 98.5\% accuracy. The results of the simulation indicated that the WOA based feature selection technique is suitable for the IDS in QNN.

Gong et al.~\cite{AA35} proposed a network attack detection scheme based on Variational Quantum Neural Network (VQNN), which consists of Variational Quantum Circuit (VQC) and classical ML strategies to simulate continuous probability distributions. Comparative experiments were conducted using the VQNN model and some classical machine learning models (such as artificial neural networks, support vector machines, K-Nearest Neighbors, Naive Bayes, and decision trees). The results indicate that the proposed IDS model based on VQNN achieved an accuracy of 97.21\%, surpassing other classical IDS models on the KDD-cup 1999 dataset they chose to use.

Akter et al.~\cite{AA26} utilized QSVM for malware classification by using penny-lane QML framework on the drebin 215 dataset. The proposed model achieved 95\% accuracy. The authors also conducted comparative experiment between QNN and NN on software supply chain attacks~\cite{AA27}. PCA is applied to reduce the dimension of the ClaMP dataset from 108 features to 16 principal components. Classical NN is directly applied to the reduced dataset. QNN model required the reduced dataset to be encoded first. They compared the f1 score, recall, precision, accuracy, and execution time. The comparative experiment indicates that QNN is slower than NN with a higher percentage of datasets. And no matter QNN or NN, both performances are below 80\% F1 score. The limitations of this research are that it is limited in the software supply chain attack and not consider other types of attack. The authors compare the performance of basic conventional neural network model with NN’s quantum version. They did not consider other further implemented NN networks.

Fioravanti et al.~\cite{AA25} proposed an open-source framework to simplify the process of simulation of quantum algorithms (fault-tolerant quantum computer). The authors conducted several experiments to observe an advantage of using quantum instead of classical machine learning algorithms, such as the comparison between Q-PCA and PCA. Their experiments show that QML may not outperform their classical counterpart in three selected public datasets, the more important point is to find a trade-off between approximation error and running time. They also found that for small datasets, they do not have an advantage in using quantum machine learning in terms of running time. As the dimensionality of the traffic dataset increases, the advantage of quantum machine learning becomes increasingly clearer.

Rahman et al.~\cite{AA29} proposed a hybrid quantum-classical algorithm called Quantum Generative Adversarial Network (QGAN). In this approach, the authors leverage the interaction between a quantum generator (ansatz) and a classical discriminator (neural network) to learn from the training NSL-KDD dataset. The training process involves adjusting the parameters of the quantum generator until it can produce output states that accurately represent the target distribution. Utilizing a quantum generator proves to be an effective means of generating samples that adhere to complex probability distributions, a task that is challenging to model using classical methods. The authors have introduced a novel perspective that integrates hybrid quantum machine learning with network intrusion detection. 

Gouveia et al.~\cite{AA30} proposed a QSVM model with auto-encoder model that uses a quantum kernel estimator and optimizer. A comparable experiment between QSVM and SVM is conducted under NSK-KDD and UNSW-NB 15 datasets. The experimental results demonstrate that QSVM performs comparably to SVM.

Thirumalairaj et.~\cite{AA31} proposed a Perimeter Intrusion Detection system with a multilayer perceptron (PID-MLP) to specify an adaptive feature that shows any number of layers can be reduced to a two-layer input-output mode. After that, a novel quantum classifier technique is used to perform the final classification. The comparative experiments with CICIDS dataset proved that the selected algorithms (PID/PID with MLP) plus a final Quantum classifier layer outperforms PID/PID with MLP without Quantum classifier. 

Mercaldo et al.~\cite{AA32} provided a comparison between five state of-the-art CNN models (i.e., AlexNet, MobileNet, EfficientNet, VGG16, and VGG19), one network developed by the authors (called Standard-CNN), and two quantum models (i.e., a hybrid quantum model and a fully quantum neural network) to classify malware. However, the authors use different image sizes, batch numbers, epoch numbers, and learning rate to do the comparison, which is unfair to certain levels. The proposed QNN and hybrid QCNN models do not outperform the classical ML models.

QSVM and QCNN as concurrent methods are discussed and evaluated in Kalinin et al.~\cite{AA33}~\cite{AA34}. The quantum version models are compared to the conventional intrusion detectors running on the classical computer with a private dataset. The comparison between ML classifiers and QML classifiers on large stream datasets has revealed the significant superiority of the quantum approach. We have summarized the above state-of-the-art (SOTA) in Table I with their proposed quantum models, datasets, comparison among different quantum models, and comparison between quantum model and classical model.

\begin{table*}[]
\centering
\caption{The summary of SOTA researches}
\begin{tabular}{|c|c|c|c|c|c|c|}
\hline
{  No.} & {  Paper} & {  Year} & {  Quantum Model} & {  Datasets} & {\begin{tabular}[c]{@{}c@{}} Comparison among \\different quantum models\end{tabular}} & {\begin{tabular}[c]{@{}c@{}} Comparison between \\quantum and classical\end{tabular}} \\ \hline
1 & Kadry et al {[}18{]} & 2023 & QNN & KDDCUP99 & NO & YES \\ \hline
2 & Akter et al {[}16{]} & 2023 & QSVM & drebin215 & NO & NO \\ \hline
3 & Rahman et al. {[}19{]} & 2023 & QGAN & NSL-KDD & NO & NO\\ \hline
4 & Kalinin et al. {[}23{]}{[}24{]} & 2023 & QSVM, QCNN & Private Dataset & YES & YES \\ \hline
5 & Akter et al {[}17{]} & 2022 & QNN & ClaMP & NO & YES \\ \hline
6 & Fioravanti et al. {[}15{]} & 2022 & QPCA & KDDCUP99, CICIDS-17, DART-NET & YES & YES \\ \hline
7 & Gong et al. {[}25{]} & 2022 & QNN & KDDCUP99 & NO & YES \\ \hline
8 & Mercaldo et al. {[}22{]} & 2022 & QCNN & Malware Software Dataset & YES & YES \\ \hline
9 & Payares et al. {[}26{]} & 2021 & QSVM,QCNN & CIC-DDoS2019 & YES & NO \\ \hline
10 & Thirumalairaj et. {[}21{]} & 2020 & QMLP & CICIDS-17 & NO & YES \\ \hline
11 & Gouveia et al. {[}20{]} & 2020 & QSVM & NSL-KDD, UNSW NB 15 & NO & YES \\ \hline
\end{tabular}
\end{table*}

Current research in this area has some limitations. Firstly, most studies have utilized existing conventional built-in QML structures from several open-sourse QML platforms while the exploration of further optimization and enhancement of these QML is limited~\cite{AA37}. There are also insufficient comparative experiments between the designed QML models and their classical ML counterparts. More often, in some studies, QMLs are compared with ML models that have low structural relevance, such as comparing a quantum CNN model with EfficientNet, rather than conducting corresponding comparisons with CNN structures~\cite{AA32}. The generalization and robustness of QML for new emerging types of traffic data are not considered as well. Secondly, many of these studies utilize outdated datasets. For example, older datasets (e.g., the KDD dataset) do not adequately represent the characteristics of contemporary network traffic. In our study, we designed five different quantum CNN models based on the underlying CNN model, utilizing custom quantum circuits, bespoke quantum layers, and various architectures of QCNN. One of the latest datasets, 5G-NIDD, is selected to conduct a comprehensive experimental analysis of the QCNN models with the corresponding CNN models. The different attack classes of this dataset are also applied to perform unseen traffic attack detection evaluation, to compare QML and Classical ML performance.

\section{\bf Methodology}
We conduct an in-depth exploration of the structure and principles of QCNN. Our plan includes designing various types of QCNN models. Additionally, we will analyze and investigate the impact of different proportions and architectures of quantum layers in hybrid QCNN on the detection performance.

\subsection{\bf \textbf{Standard Convolutional Neural Network:}}
Convolutional neural network is a deep learning model. It is mainly used for tasks such as image recognition, computer vision, and image processing. It has also been successfully applied in the field of malicious traffic monitoring and has performed well~\cite{AA39}~\cite{AA40}~\cite{AA41}. The CNN model can accurately detect image-like arrays generated by network traffic data processing. CNN usually consists of a series of different image processing layers; in each layer, the previous layer produces an intermediate array of pixels, i.e., a feature map~\cite{AA38}. CNN is composed of different types of layers, including convolution layers and pooling layers forming a convolution block, as well as fully connected layers forming a fully connected block. The convolution block, comprising alternating convolution and pooling layers, forms an input feature vector for the fully connected block. CNN gradually extracts the abstract features of the input data by stacking these different types of layers and finally outputs a high-level representation of the input.

\textbf{Baseline Model: Standard CNN model (Basic CNN)}

The initial model is a classical CNN. We keep it simple for comparison with its quantum counterparts which ensures that extensive classical parameters do not skew the comparison. Despite its simplicity, it is crucial that the classical CNN achieves comparable standards of accuracy to its quantum counterparts, ensuring a fair evaluation. This comparative accuracy will be detailed in the experimental section. The CNN's convolution block comprises two convolution layers, each followed by max pooling. The fully connected block consists of one internal dense layer and a final prediction layer for making the ultimate prediction. The structure of the model is shown in Figure 1. More channels in CNN can extract more features. Our dataset contains 28 initial features, thus, we choose to use 32 channels to extract more internal features. And 32 channels are not large, thus, it will not consume lots of computation resources and running time. The following QCNN models are designed based on this standard CNN model.

\begin{figure}[h]
\centerline{\includegraphics[width=21pc]{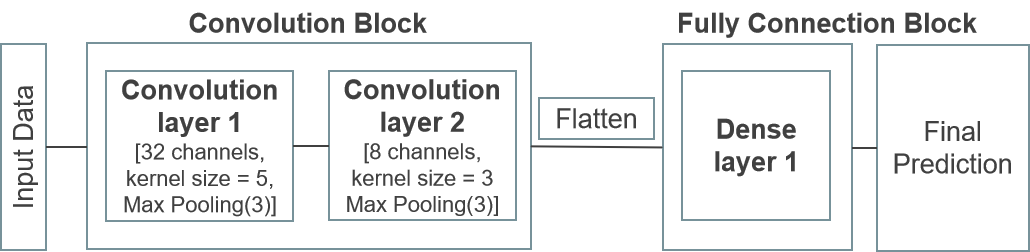}}
\caption{The structure of the baseline CNN model}
\end{figure}

\subsection{\bf \textbf{Quantum Computing and Machine Learning}}

Quantum Computing is a computing method that processes information based on the principles of quantum mechanics. Quantum circuit is a model to implement quantum computing which consists of quantum gates (operate on qubits) and measurement operations. Qubit is considered as the basic unit for constructing a quantum circuit. It is used to represent and process high-dimensional data. Unlike a classical bit that has 0 and 1 two states to process information, the corresponding qubits in quantum computing possess additional properties, such as superposition and entanglement.

Quantum entanglement refers to the fact that the state of one qubit can depend on the states of other qubits. Superposition, on the other hand, refers to a qubit that can be in a linear combination of \(|0\rangle\) and \(|1\rangle\) states simultaneously. The state of a qubit can be expressed as:

\(|\psi\rangle = \alpha|0\rangle + \beta|1\rangle\)

We can also obtain a visual representation of a qubit on the surface of the Bloch sphere:

\(|\psi\rangle = \cos\left(\frac{\theta}{2}\right)|0\rangle + e^{i\phi} \sin\left(\frac{\theta}{2}\right)|1\rangle\)

It is easy to analyze various operations and transformations of qubits by using Bloch sphere. Thus, complex problems in classical computing can be presented in new ways in multi-dimensional computational spaces which are created by the superposition and entanglement of qubits. QML relies on the properties of qubits and quantum circuits to enhance classical ML algorithms, making them more effective in solving complex problems. QML can be quantum versions of classical ML or quantum-classical hybrid algorithms. Currently, QML has become a new trend in quantum computing applications.

\subsection{\bf \textbf{Quantum Convolutional Neural Network:}}
QML integrates quantum algorithms into the field of classical machine learning models. Hybrid-QCNN consists of functional blocks based on both quantum and classical neural networks. It can be interpreted as incorporating a certain number of quantum layers into a classical convolutional neural network. Different quantum layers can be designed to have various functions (e.g., data quantum embedding, quantum convolution, or quantum pooling). The overall objective is to use transformations in quantum circuits to simulate the behavior of quantum computers on classical computers. In this paper, we designed five QCNN structures based on classical CNN models to investigate the impact of quantum embedding methods, the complexity of quantum layers, and the positions of quantum layers in the QCNN model by using the Pennylane software framework~\cite{AA42}.

\textbf{QCNN1: QCNN with Angle Embedding (QCNNAnE)}

The first hybrid QCNN model introduces quantum layers with angle embedding applied to the fully connected block, creating a hybrid quantum fully connected block. Angle embedding refers to a technique for representing classical data as quantum states. It involves encoding the information of features or classical data points into the angles of quantum states, thereby mapping them onto quantum states. It can encode N features or data points into the rotation angles of n qubits, where $N \leq n$. Rotation gates $R_x(\theta)$, $R_x(\theta)$, and $R_x(\theta)$ in a quantum circuit encode classical feature information into rotation angles to achieve a quantum state representation:

\(|\psi\rangle = R_x(\theta) |0\rangle\)

\(|\psi\rangle = R_y(\theta) |0\rangle\)

\(|\psi\rangle = R_z(\theta) |0\rangle\)

The angle \(\theta\) is determined based on the value of classical feature input and \(R_x(\theta)\) means the rotation gate rotates the state around D-axis (x-axis, y-axis, z-axis) on the Bloch sphere by an angle \(\theta\). 

Classical input data undergoes angle embedding to transition into a quantum state, subsequently passing to the quantum layer. In order to facilitate a fair comparison of the various QCNN models we designed, we ensured a consistent approach in the design of the quantum circuits. The circuit will always consist of one-parameter single qubit rotations on each qubit, followed by a closed chain of Controlled NOT (CNOT) gates. CNOT gates link each qubit with its adjacent neighbor, including the last qubit, which is treated as a neighbor to the first qubit. This type of quantum circuit structure is called a Parameterized Quantum Circuit (PQC)~\cite{AA23}. It is a widely used quantum circuit structure in quantum machine learning. This circuit design ensures the efficient entanglement of qubits. It also promotes the exploration of the full range of quantum states.

CNOT gate is a quantum gate that has two qubits. The first qubit refers to as the control qubit, \(|c\rangle\), and the other qubit refers to as the target qubit,  \(|t\rangle\). The operation of the CNOT gate can be introduced as follows:

If \(|c\rangle = |0\rangle, \quad |t\rangle\) \text{ remains unchanged.}

If \(|c\rangle = |1\rangle, \quad\) \text{NOT gate applied to } \(|t\rangle\). 

The quantum circuit design manner is shown in Figure 2. In Figure 2, there are 4 qubits in the circuit and the chain of gate connects every qubit with its neighbor. These 4 qubits can encode 4 features onto quantum state. We can use classical layers (convolution, pooling, and fully connected layers) to reduce the initial number of feature set to four features that match the requirements of this quantum circuit. The structure of the QCNNAnE can be found in Figure 3.

\begin{figure}[h]
\centerline{\includegraphics[width=21pc]{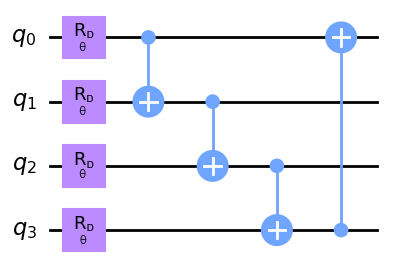}}
\caption{The structure of quantum circuit design manner. D refers to the rotation dimension. $\theta$ refers to the angle value}
\end{figure}

\begin{figure}[h]
\centerline{\includegraphics[width=21pc]{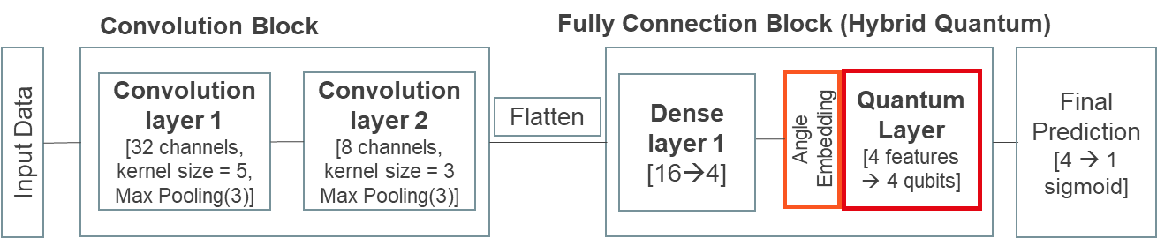}}
\caption{The structure of QCNNAnE}
\end{figure}

In Figure 3, we illustrate the process using a sample input with 28 features. As an illustrative example, upon traversing the convolution layers and dense layer 1, all features of the input sample are encoded into a quantum state. Following this, the input data, now represented in a quantum state, is directed to the quantum layer. Several quantum operations are applied to process and extract information from the quantum state. Subsequently, the final expectation values from the quantum layer are measured and incorporated into the final prediction.

\textbf{QCNN2: QCNN with Amplitude Embedding (QCNNAmE)}

The difference between QCNNAmE and QCNNAnE is the embedding method used in converting classical input data to quantum state. In QCNNAmE, amplitude embedding method is applied in the QCNN model. The amplitude embedding is employed to encode classical data to probability amplitudes of quantum state. A quantum state can exist simultaneously in a superposition of multiple states, with each state associated with an amplitude. By manipulating these amplitudes, quantum algorithms can perform various computations more efficiently than classical algorithms. A normalized classical data x with N-dimension can be represented by the amplitudes of an n-qubits quantum state \(|\psi\rangle\):

\(|\psi\rangle = \sum_{i=1}^{N} x_i|i\rangle\), where \(x_i\) means the i-th data, and \(|i\rangle\) refers the i-th computational basis state.

The structure of QCNNAmE is shown in Figure 4, it closely resembles the second model, with the quantum layer integrated into the fully connected block. The key difference is the data embedding method changes from angle embedding to amplitude embedding. The disparity between angle embedding and amplitude embedding lies in the qubits requirements for representing a data point. Amplitude embedding demands a smaller number of qubits to represent a data point, whereas angle embedding requires the same number of qubits as the features of the data point in QCNN model design. We conduct a comparative analysis between QCNNAmE and QCNNAnE to study the effect of different quantum embedding methods in QCNN models. 

Due to the properties of amplitude embedding, we can utilize n qubits to encode \(2^n\) features. Consequently, the model eliminates the need for Dense Layer 1 to reduce dimensions like QCNNAnE. This is beneficial for encoding large datasets onto quantum states because it reduces the resource overhead. However, it may also reduce the robustness to noise and errors in some data because it relies on the amplitude rather than the relative phase angle.

\begin{figure}[h]
\centerline{\includegraphics[width=21pc]{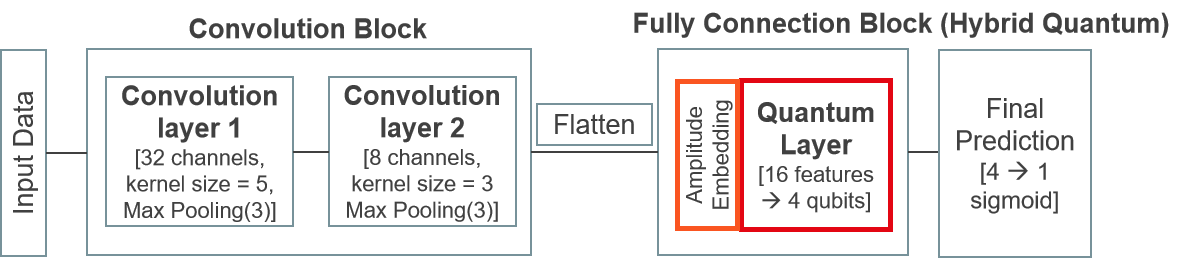}}
\caption{The structure of QCNNAmE}
\end{figure}

\textbf{QCNN3: QCNN with Multi-quantum Layers in Parallel (QCNNMlayer)}

The structure of QCNNMlayer is shown in Figure 5. QCNNMlayer is designed with the aim of augmenting the quantum structure's proportion and complexity within the hybrid QCNN model. To achieve this, we design an architecture that incorporates a classical convolution block alongside multiple parallel quantum layers, effectively combining them. This architecture draws inspiration from hybrid convolution function with multiple quantum filters~\cite{AA41}. The difference is that the hybrid convolution function performs quantum convolution by replacing parallel classical convolutional filters with quantum filters. In contrast, parallel quantum layers of QCNNMlayer play the role of a fully connected layer, which are not responsible for the convolution of the input data.

Following the classical convolution of the data, it is flattened into an array containing 16 features. Subsequently, the model partitions these 16 features into four sets, with each set forwarded to a dedicated quantum layer. The final measured expectation values are then combined and passed on to the final prediction layer.

\begin{figure}[h]
\centerline{\includegraphics[width=21pc]{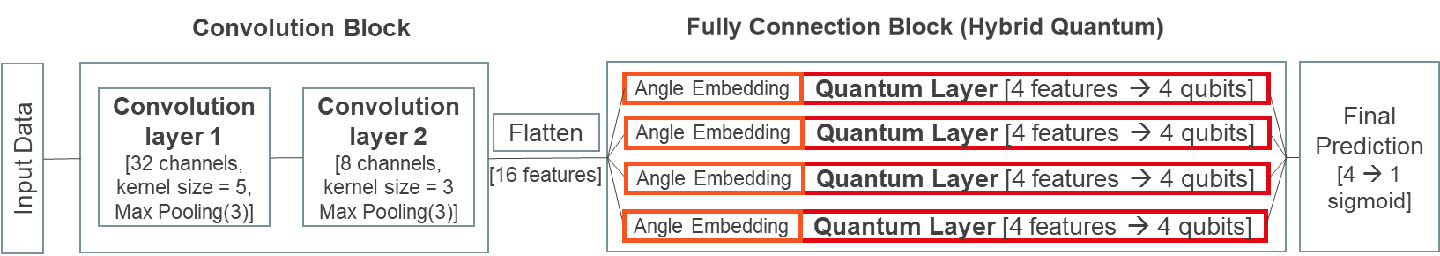}}
\caption{The structure of QCNNMlayer}
\end{figure}

\textbf{QCNN4: QCNN under Quantum Convolution (Quan-ConvCNN)}

Starting from Quan-ConvCNN, we shift our focus to introducing quantum computing into the convolution block. Our objective is to make quantum layers emulate the effects of classical convolution. Thus, We decided to optimize the quantum computing process by systematically reducing the measurements of qubits. This optimization involves performing operations on each qubit until reaching a predefined stage. At this point, we decided to selectively disregard specific qubits in specified layers of the quantum circuit. This strategic approach aims to increase the efficiency of quantum computation while ensuring that computational resources are utilized appropriately. It is these layers where we stop performing operations on certain qubits that we call our ‘convolution layer’. The structure of the quantum convolution circuit is shown in Figure 6.

\begin{figure}[h]
\centerline{\includegraphics[width=18pc]{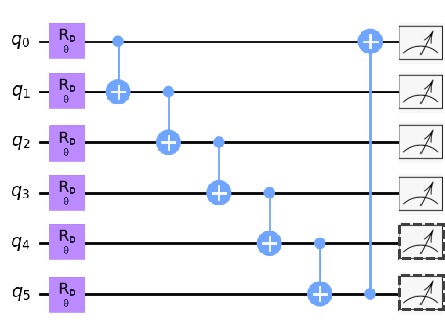}}
\caption{The structure of quantum circuit design manner in Quan\-ConvCNN}
\end{figure}

In Figure 6, in order to achieve a similar performance of classical convolution and pooling (discard some features to reduce input feature dimension), we randomly disregard two qubits (in the Figure 6 example, q4 and q5 are disregarded) and only consider outputs of remaining 4 qubits. This approach incorporates quantum computing into the convolution block, expanding the quantum influence in our models.

\begin{figure}[h]
\centerline{\includegraphics[width=21pc]{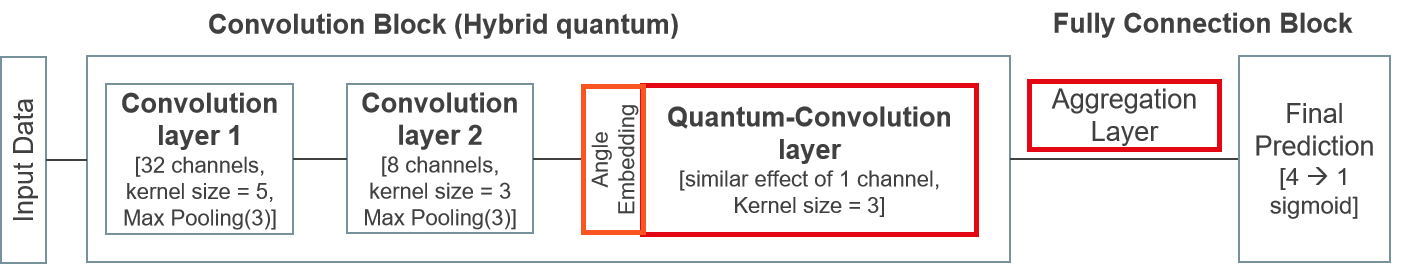}}
\caption{The structure of Quan-ConvCNN}
\end{figure}

Figure 7 illustrates the structure of Quan-ConvCNN. With this quantum-convolution layer, it can achieve a similar performance of 1 channel filter with kernel size =3. After aggregating expectation values from quantum convolution layer, the output data can be passed to the final prediction layer directly. Algorithm 1 illustrate the training process of Quan-ConvCNN.

\begin{algorithm}
\caption{The algorithm of quantum convolution in Quan-ConvCNN model}
\begin{algorithmic}[1]
\Require 
\State $w(\text{weight}) \leftarrow \text{define weights}$
\State $b(\text{bias}) \leftarrow \text{define bias}$
\State $Qbits(\text{no. of qubits}) \leftarrow \text{define the number of qubits}$
\State $x(\text{input data}) \leftarrow \text{Reshape} \leftarrow [\text{Dimension}, \text{channels}]$
\Function{QCNN}{$x$}
    \State $conv1 \leftarrow \text{ReLU}(\text{Conv1D}(x, w[0] + b[0]))$
    \State $conv1 \leftarrow \text{Max\_Pooling}(\text{Conv1D}(conv1))$
    \State $conv2 \leftarrow \text{ReLU}(\text{Conv1D}(conv1, w[1] + b[1]))$
    \State $conv2 \leftarrow \text{MaxPooling}(\text{Conv1D}(conv2))$
    \State $qconv1 \leftarrow \text{Angle\_embedding}(conv2)$
    \State $quantum\_eval \leftarrow \text{qconvolution}(Qbits, qconv1)$
    \If{$\text{qubit is not a disregarded qubit}$}
        \State $\text{measure\_expectation\_value}(\text{Pauli\_Z}(\text{qubit}))$
    \Else
        \State \textbf{disregard}
    \EndIf
    \State $quantum\_eval \leftarrow \text{Aggregate}(quantum\_eval)$
    \State $\text{flatten()}$
    \State $\text{fully\_connect} \leftarrow \text{ReLU}(quantum\_eval, w[3] + b[3])$
    \State \textbf{Output} $\leftarrow \text{Sigmoid}(\text{fully\_connect})$
\EndFunction
\end{algorithmic}
\end{algorithm}

\textbf{QCNN5: Alternative Strategy for Quantum Convolution Based NN (QuanvolutionNN)}

QuanvolutionNN represents an alternative strategy for achieving quantum convolution, drawing inspiration from Henderson et al.~\cite{AA43}. The key difference in this hybrid QCNN, compared to previous models, lies in the approach of embedding a small region of the input image into a quantum circuit at a time. The quantum convolution procedure can be found in Figure 8. Analogous to a classical convolution layer, each expectation value resulting from this quantum convolution is mapped to a distinct channel of a single output pixel. Figure 9 illustrates an example where 30 traffic instances pass through the quanvolution layer, generating two new channel images with halved dimensions. In essence, this quantum convolution algorithm achieves a similar effect to employing 2 channel filters with a kernel size of 1 and a pooling size of 2. This innovative approach provides a unique perspective on incorporating quantum principles into convolutional neural networks.

\begin{figure}[h]
\centerline{\includegraphics[width=21pc]{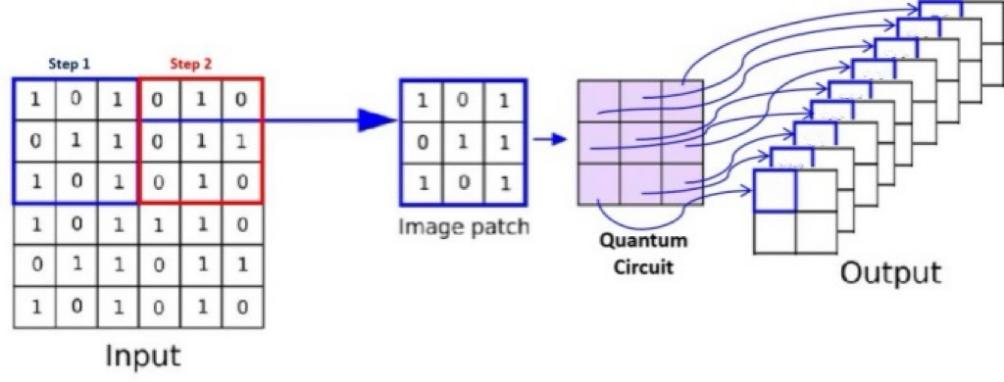}}
\caption{The procedure of quanvolution}
\end{figure}

\begin{figure}[h]
\centerline{\includegraphics[width=8pc]{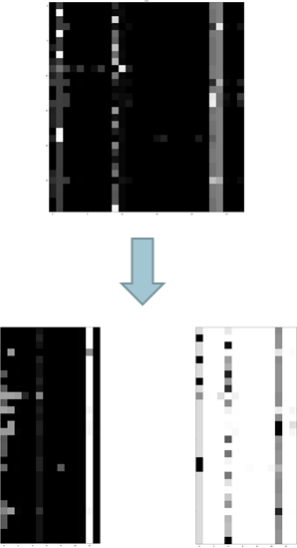}}
\caption{Example where 30 traffic instances pass through the quanvolution layer}
\end{figure}

In QuanvolutionNN, the quanvolution layer is positioned ahead of all classical convolution layers. By configuring the parameters in the quanvolution layer as non-trainable, it can be regarded as a data processing step. This approach allows for a flexible interpretation, considering the quanvolution layer as either an integral part of the learning process or as a predetermined data processing step based on the specific needs of the model. The Structure of the QuanvolutionNN model is shown in Figure 10.

\begin{figure}[h]
\centerline{\includegraphics[width=21pc]{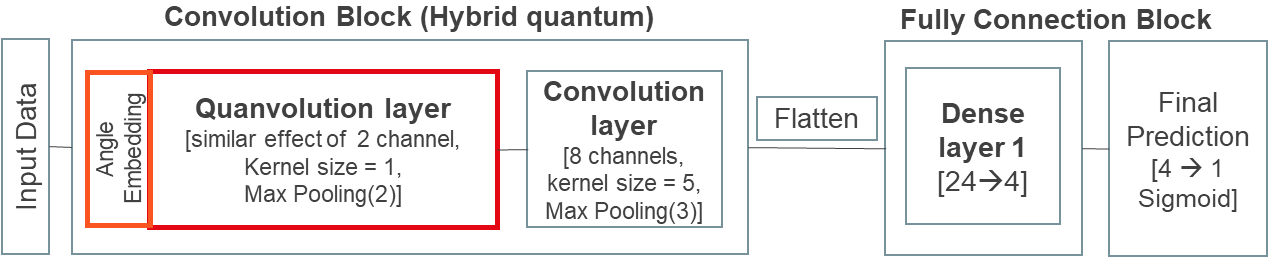}}
\caption{The structure of QuanvolutionNN}
\end{figure}

\section{\bf Experiment}

\subsection{\bf \textbf{Experiment Set-up}}

The experiment running: Intel(R) Core(TM) i7-10700K CPU @ 3.8GHz 64.0GB of RAM. In order to construct our models, Pytorch, Qiskit and Pennylane libraries for Python are used. Pennylane provides an efficient QML framework for the design of our quantum circuits and models, and seamlessly integrates into machine learning models. It also offers a user-friendly interface for quantum computers provided by industry leaders such as IBM, Google, or Microsoft~\cite{AA18}.

In this section, we conduct 3 experiments to evaluate the performance of QCNNs and perform comparison of QCNNs with classical CNN. For experiment 1, we perform the standard ML evaluation where random train and test set split from the selected dataset is applied. In experiment 2, the train set and test set are considered from different data collection positions so as to conduct an unseen incoming traffic detection. More details of unseen incoming traffic detection can be found in Section V.D Experiment 2. Experiment 3 is to filter out one specific attack type from the training set and not train on it, only using this attack type in the test set. This is aimed to test the generalization and robustness of the model.

In order to ensure the fairness of the comparative experiments, it is imperative to maintain constancy in most control variables. Key parameters such as learning rate, epoch number, and optimizer configurations are standardized across all experiments. For instance, we set the epoch number as 30 across all experiments to ensure the fairness of comparison. 30 epochs are sufficient for training the model to achieve comparable performance as well. In the domain of quantum computing, meticulous efforts are undertaken to uphold uniformity in the number of qubits and wires, either by maintaining consistency or within a small range of qubits number change. Moreover, the quantum circuit adheres consistently to a predefined design methodology, as mentioned in Section 3, irrespective of the chosen number of qubits. 

Based on the results of multiple experiments, we have observed that the performance of QCNN models is highly dependent on the randomly generated weights of the quantum layers before model training. This randomness introduces a stochastic element issue, and if the randomly generated quantum layer weights are well-suited for detection, the model exhibits excellent convergence speed and accuracy. However, if the randomly generated quantum weights are not appropriate, issues may arise in terms of detection accuracy, such as gradient vanishing or getting stuck in a local minimum. In order to solve this issue, we decide to conduct experiments 1 and 2 three times and calculate the average performance result to ensure fair experiment results. The following experiment results are the average result of each QCNN model. Each individual QCNN experiment results are listed in the Appendix. 

\subsection{\bf \textbf{Dataset and feature selection}}

As mentioned earlier, many of the datasets used in current research are outdated, such as CTU-13 and KDD-CUP 99 datasets. In this paper, we decide to utilize one of the latest publicly available datasets, 5G NIDD dataset. 5G-NIDD is a fully labeled dataset created by Samarakoon et al.~\cite{AA40} based on the Finnish 5G Test Network~\cite{AA39} (5GTN). To create the dataset, the authors selected the University of Oulu site in the 5GTN. Data collection took place on two different days and captured both attack traffic and benign traffic passing through the two base station networks. The 5G-NIDD dataset is available in various versions, including packet-based and traffic-based formats, which greatly improves the dataset's mineability. Data from each attack session from both base stations is provided separately in pcapng format. With the removal of the Generic Packet Radio Service Tunneling Protocol (GTP) layer, the same files are available in packet-based pcapng, argus, and csv feature formats. The traffic-based files after the removal of the GTP layer are combined to form a single file containing all attacks from both base stations as well. Table II is the statistical summary of the 5G NIDD dataset.

\begin{table}[h]
\centering
\caption{The statistical analysis of 5GNIDD dataset}
\begin{tabular}{|c|c|c|c|c|}
\hline
\multicolumn{1}{|c|}{No.} & \multicolumn{1}{c|}{\begin{tabular}[c]{@{}c@{}}Type of\\ Traffic\end{tabular}} & \multicolumn{1}{c|}{\begin{tabular}[c]{@{}c@{}}Base\\ Station\end{tabular}} & \multicolumn{1}{c|}{\begin{tabular}[c]{@{}c@{}}No. of \\ Flow Session\end{tabular}} & \multicolumn{1}{c|}{\begin{tabular}[c]{@{}c@{}}Malicious\\ or\\ Benign\end{tabular}} \\ \hline
\multirow{2}{*}{1} & Legitimate traffic        & BS1          & 406959              & Benign           \\ \cline{3-5} 
                   &                                 & BS2          & 70778               & Benign           \\ \hline
\multirow{2}{*}{2} & UDPFlood       & BS1          & 175811              & Malicious        \\ \cline{3-5} 
                   &                                 & BS2          & 281529              & Malicious        \\ \hline
\multirow{2}{*}{3} & HTTPFlood      & BS1          & 76121               & Malicious        \\ \cline{3-5} 
                   &                                 & BS2          & 64691               & Malicious        \\ \hline
\multirow{2}{*}{4} & SlowrateDoS    & BS1          & 36092               & Malicious        \\ \cline{3-5} 
                   &                                 & BS2          & 37032               & Malicious        \\ \hline
\multirow{2}{*}{5} & TCPConnectScan & BS1          & 10022               & Malicious        \\ \cline{3-5} 
                   &                                 & BS2          & 10030               & Malicious        \\ \hline
\multirow{2}{*}{6} & SYNScan        & BS1          & 10019               & Malicious        \\ \cline{3-5} 
                   &                                 & BS2          & 10024               & Malicious        \\ \hline
\multirow{2}{*}{7} & UDPScan        & BS1          & 7887                & Malicious        \\ \cline{3-5} 
                   &                                 & BS2          & 8019                & Malicious        \\ \hline
\multirow{2}{*}{8} & SYNFlood       & BS1          & 4792                & Malicious        \\ \cline{3-5} 
                   &                                 & BS2          & 4929                & Malicious        \\ \hline
\multirow{2}{*}{9} & ICMPFlood      & BS1          & 613                 & Malicious        \\ \cline{3-5} 
                   &                                 & BS2          & 542                 & Malicious        \\ \hline
\end{tabular}
\end{table}

Table III is the feature set we selected in the experiment. There are 28 features selected based on Pearson correlation and expert selection. We keep selected features consistent in all experiments. After feature selection, the numerical ranges of different features may vary greatly. Thus, data normalization needs to be applied. Data normalization can normalize the traffic data to a range of zero to one to reduce data redundancy. The normalized dataset enhances the integrity and efficiency of model training.

\begin{table}[h]
\centering
\caption{Feature Set Selection}
\begin{tabular}{|c|l|c|l|}
\hline
No. & Feature Name             & No. & Feature Name          \\ \hline
1   & Dur                      & 2   & STtl                  \\ \hline
3   & dTtl                     & 4   & TotPkts               \\ \hline
5   & SrcPkts                   & 6   & DstPkts               \\ \hline
7   & TotBytes                  & 8   & SrcBytes              \\ \hline
9   & DstBytes                  & 10  & Offset                \\ \hline
11  & sMeanpktSz               & 12  & dMeanPktSz            \\ \hline
13  & Load                     & 14  & SrcLoad               \\ \hline
15  & DstLoad                   & 16  & Loss                  \\ \hline
17  & SrcLoss                   & 18  & DstLoss               \\ \hline
19  & SrcWin                    & 20  & DstWin                \\ \hline
21  & TcpRtt                   & 22  & SynAck                \\ \hline
23  & AckDat                   & 24  & Rate                  \\ \hline
25  & SrcRate                  & 26  & DstRate               \\ \hline
27  & SrcTCPBase               & 28  & DstTCPBase            \\ \hline
\end{tabular}
\end{table}

\subsection{\bf \textbf{Experiment 1:}}

Experiment 1 is designed to validate the performance of various QCNN models and classical CNN counterpart in traditional offline ML training and testing. This is also a common ML experiment in the field of traffic detection research. Without considering that the trained model would be applied to the detection of unknown traffic type or new incoming network traffic, the whole dataset is partitioned into 851,123 train set and 364,747 test sets (7:3 ratio). Given the substantial variation in numerical ranges among different features, data normalization needs to be applied. The utilization of a normalized dataset contributes to the integrity and efficiency of model training. The average Experiment 1 performance results of each model are recorded in Table IV and plotted in Figure 11. Each individual experiment result is recorded as well in the Appendix.

\begin{table}[]
\centering
\caption{Average Accuracy Performance of QCNN \& CNN Models in Experiment 1}
\begin{tabular}{|c|c|}
\hline
Detection Model & Accuracy (average) \\ \hline
CNN             & 99.61\%            \\ \hline
QCNNAnE         & 99.62\%            \\ \hline
QCNNAmE         & 99.65\%            \\ \hline
QCNNMlayer      & 98.89\%            \\ \hline
Quan-ConvCNN    & 99.71\%            \\ \hline
QuanvolutionNN  & 98.67\%            \\ \hline
\end{tabular}
\end{table}

\begin{figure}[h]
\centerline{\includegraphics[width=21pc]{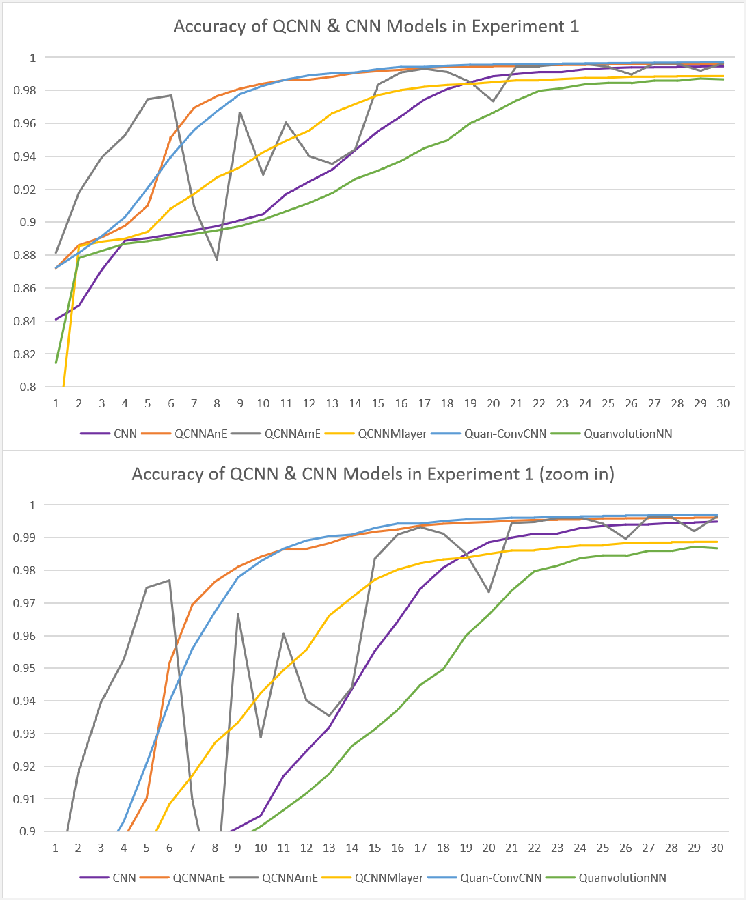}}
\caption{Average Accuracy Performance of QCNN \& CNN Models in Experiment 1}
\end{figure}

Based on the experimental results, we can find that most Hybrid QCNN models outperform the Classical CNN. According to the above graphs drawn in 30 epochs, we find that QCNNAmE has the fastest convergence speed, reaching over 97.5\% accuracy in the 5th epoch, while classical CNN reaches over 98\% only in the 13th epoch. QCNNAnE and Quan-ConvCNN have a slower convergence speed than QCNNAmE, but are still better than classical CNN, and the model improvement during the training process is smooth without violent oscillations.

It is worth noting that although QCNNAmE converges faster and achieves the second highest accuracy, its oscillations during the training process are very obvious, for example, at the 6th epoch, its accuracy drops from over 97.8\% to around 92\%. This may be due to the fact that we embedded 16 features into only 4 qubits, which caused the model to lose some useful information, resulting in significant oscillations when updating the model. Quan-ConvCNN achieves the highest accuracy without significant oscillations, but a relatively slower convergence speed than QCNNAmE. 

In addition, our QCNNMlayer has a worse convergence performance, its experimental results do not exceed the classical CNN. The reason may be that the number of parameters in quantum layers of QCNNMlayer is larger than that of other QCNN models. For the QCNNMlayer model, following the classical convolution of the data, it is flattened into an array containing 16 features. Subsequently, the model partitions these 16 features into four sets, with each set forwarded to a dedicated quantum layer with 4 qubits. Thus, there are total 16 qubits are utilized in this model. Thus, there are more complicated quantum weights to be computed and updated. This leads to the slow convergence of the model and needs to be trained more than 30 epochs. The detailed architecture and qubit number used for QCNNMlayer are described in Section IV.B. For QuanvolutionNN, the experiment results indicate that the process of halving the dimensions of images using quanvolution method leads to a significant loss of data information. This results in both slower convergence speed and inferior accuracy.

In Quan-ConvCNN, the quantum layer is designed to perform convolution that achieves a higher result than other QCNN models, such as using quantum layer as fully connected layer in the hybrid CNN model. Quan-ConvCNN also shows a relatively better performance convergence speed and stability than other QCNN models. Such results indicate that designing quantum layer to perform convolution in Quan-ConvCNN algorithm may be better than designing quantum layer as fully connected layer or quanvolution layer of QuanvolutionNN. 

We also found that QCNN model easily falls into the local minimum point and without any further improvement in accuracy performance (i.e., the model stack at ~91.5\% and does not improve anymore.) in experiments. Such unsuccessful experiments are disregarded.

\subsection{\bf \textbf{Experiment 2:}}

Experiment 2 is purposed to perform unseen incoming traffic data detection so as to evaluate the robustness of models. After model training, the model will test with new traffic data from another data base station but with identical malicious traffic types. For the selected 5G NIDD dataset, it is collected from two base stations. 728,316 traffic sessions are collected from Base Station 1 and 487,574 traffic sessions are collected from Base Station 2. Due to differences in location, traffic purposes, and collection time, traffic data collected at two base stations may exhibit different distributions even though the types of traffic are identical. Thus, one base station's data of the 5G NIDD dataset is selected as the train set and the other one is selected as test set. We use this train and test set to simulate new incoming unseen data detection situation to test the robustness of the model.

Experiment 2 is divided into 2 sub-experiments. Sub-experiment 2.1 is to use base station 1 as the train set and base station 2 as the test set. Sub-experiment 2.2 is to exchange both data base stations.

For the sub-experiment 2.1, performance results are shown in Figure 12 and Table V. Firstly, compared with experiment 1 results, the experiment results in experiment 2 indicate that All models' performance dropped from above 98\% to around 78\% due to such unseen traffic from different base station. This further proves that two base stations have different traffic patterns.

The performance results of all models in sub-experiment 2.1 are close. The Quan-ConvCNN achieves the highest 78.93\% accuracy performance again. All QCNN models outperform the classical CNN model. Therefore, except for QCNNMlayer, the performance of other models remains consistent with that of Experiment 1. In sub-experiment 2.1, the performance of QCNNMlayer exceeds that of classical CNN, indicating that it has better robustness. QCNNAmE achieved the second highest performance, but it still has a large oscillation in this experiment. This experiment proves again that QCNNAmE which uses fewer qubits to embed more information would make the model training process unstable. 

\begin{table}[]
\centering
\caption{Average Accuracy Performance of QCNN \& CNN Models in Sub-Experiment 2.1}
\begin{tabular}{|c|c|}
\hline
Detection Model & Accuracy (average) \\ \hline
CNN             & 78.23\%            \\ \hline
QCNNAnE         & 78.30\%            \\ \hline
QCNNAmE         & 78.81\%            \\ \hline
QCNNMlayer      & 78.54\%            \\ \hline
Quan-ConvCNN    & 78.93\%            \\ \hline
QuanvolutionNN  & 78.66\%            \\ \hline
\end{tabular}
\end{table}

\begin{figure}[h]
\centerline{\includegraphics[width=21pc]{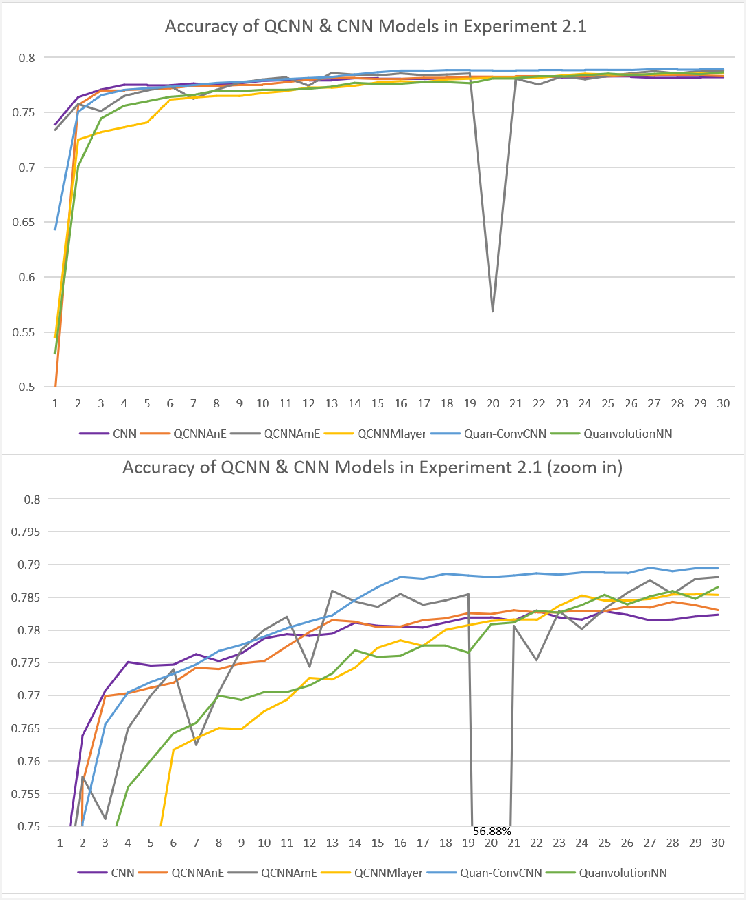}}
\caption{Average Accuracy Performance of QCNN \& CNN Models in Sub-Experiment 2.1}
\end{figure}

For the sub-experiment 2.2, this experiment is slightly different from the above sub-experiment 2.1. The reason is the size of data in base station 2 is smaller than base station 1, especially after down sampling the base station 2 train set for data balance. This experiment can be viewed as we are using a small size dataset to train the hybrid quantum model and test it with a large size dataset that contains much more unseen traffic data distribution. the performance results are shown in Figure 13 and Table VI

\begin{table}[]
\centering
\caption{Average Accuracy Performance of QCNN \& CNN Models in Sub-Experiment 2.2}
\begin{tabular}{|c|c|}
\hline
Detection Model & Accuracy (average) \\ \hline
CNN             & 74.28\%            \\ \hline
QCNNAnE         & 59.36\%            \\ \hline
QCNNAmE         & 81.57\%            \\ \hline
QCNNMlayer      & 69.70\%            \\ \hline
Quan-ConvCNN    & 64.07\%            \\ \hline
QuanvolutionNN  & 57.18\%            \\ \hline
\end{tabular}
\end{table}

\begin{figure}[h]
\centerline{\includegraphics[width=18pc]{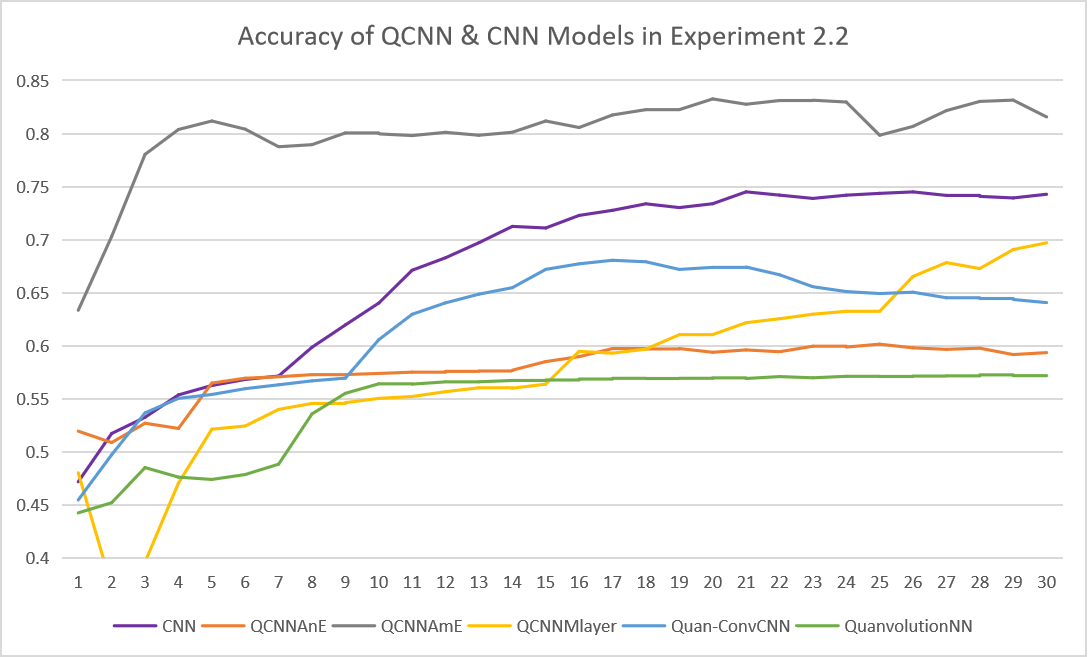}}
\caption{Average Accuracy Performance of QCNN \& CNN Models in Sub-Experiment 2.2}
\end{figure}

From the above experimental results, in the case of a small training dataset, all QCNN models using angle embedding exhibit inferior performance compared to the classical CNN model. For instance, the previously best-performing Quan-ConvCNN model, when faced with a small dataset, gets a 10.21\% lower performance than the classical CNN. QCNNAmE using amplitude embedding continues to demonstrate the fastest convergence speed and achieves the highest accuracy of 81.57\%. This may suggest that in the case of a small dataset, amplitude embedding has a higher probability of capturing important features compared to angle embedding. However, it is noteworthy that QCNNAmE exhibits the most pronounced oscillations during the training process. Its performance variations are also significantly higher than other models. 

Based on the results of sub-experiments 2.1 and 2.2, in the case of non-small dataset sizes, QCNNs demonstrate better robustness compared to classical CNNs. However, in terms of training process stability, classical CNNs typically feature a stable training process for classical data. This stability can lead to faster convergence and better robustness performance on small size datasets. In contrast, QCNN models may be more affected by insufficient training data.

\subsection{\bf \textbf{Experiment 3:}}

The third experiment is to process the data based on the type of malicious traffic. The 5G NIDD dataset contains 8 types of malicious traffic, we select one type as the test set and train with all remaining data so as to test the generalization of the model. Generalization of a detection model is crucial because it reflects how well the detection model can perform on unseen traffic types out of the training data, such as zero-day attack detection. The statistical analysis of each attack type in the 5G-NIDD dataset is shown in Table VII:

\begin{table}[]
\centering
\caption{The statistical summary of each attack types in 5G NIDD dataset}
\begin{tabular}{|c|c|c|c|}
\hline
No. & Attack Type    & Sample Size & Percentage \\ \hline
1   & UDPFlood       & 457340      & 37.61\%    \\ \hline
2   & HTTPFlood      & 140812      & 11.58\%    \\ \hline
3   & SlowrateDoS    & 73124       & 6.01\%     \\ \hline
4   & TCPConnectScan & 20052       & 1.65\%     \\ \hline
5   & SYNScan        & 20043       & 1.65\%     \\ \hline
6   & UDP Scan       & 15906       & 1.31\%     \\ \hline
7   & SYNFlood       & 9721        & 0.8\%      \\ \hline
8   & ICMPFlood      & 1155        & 0.09\%     \\ \hline
\end{tabular}
\end{table}

Based on the attack type percentage in Table VII, UDP Scan, ICMPFlood, SYNFlood, TCPConnectScan, and SYNScan in the dataset that have a very small proportion (less than 2\%). Therefore, in order to simplify the experiment, we decided to combine all attack types that make up less than 2\% of the total dataset as one test set to conduct model evaluation. This means that the model will train with benign, HTTPFlood, SlowrateDos, and UDP Flood traffic data and test the remaining attack types at one time. The new attack category table is shown in Table VIII.

\begin{table}[]
\centering
\caption{The statistical summary of new attack categories}
\begin{tabular}{|c|c|c|c|}
\hline
No. & Attack Type    & Sample Size & Percentage \\ \hline
1   & UDPFlood       & 457340      & 37.61\%    \\ \hline
2   & HTTPFlood      & 140812      & 11.58\%    \\ \hline
3   & SlowrateDoS    & 73124       & 6.01\%     \\ \hline
4   & Remain Type       & 66877       & 5.50\%     \\ \hline
\end{tabular}
\end{table}

In addition, in this experiment, QCNNMlayer was excluded from consideration. This decision was motivated by the structure of QCNNMlayer is similar to that of QCNNAnE but is more complex and requires much longer training time than QCNNAnE. Furthermore, the experimental performance of QCNNMlayer demonstrated limited improvement over QCNNAnE, and in some instances, it even exhibited inferior results such as experiment 2.1.

Due to the nature of Experiment 3, two performance results will be recorded. One is the performance of the model after fixed 30 epochs. The other one is the trade-off point of train set and test set accuracy. The decision to set the model training epoch at 30 was determined through experimentation. We conducted tests with epoch values of 30, 50, and 100, observing that within the first 30 epochs, the model sufficiently identified an optimal trade-off point. The criteria of the trade-off point are based on the intersection point of train set accuracy and test set accuracy (in the condition that the model initially predicts all test set as malicious). If there is no intersection point, we first calculate the absolute difference value in between train and test set accuracy. Then the epoch with the minimum difference value will be considered as the trade off point. Figure 14 is an illustrative example of this observation. The Figure 14 experiment is the QCNNAnE trained in the Remain Type test. 

\begin{figure}[h]
\centerline{\includegraphics[width=21pc]{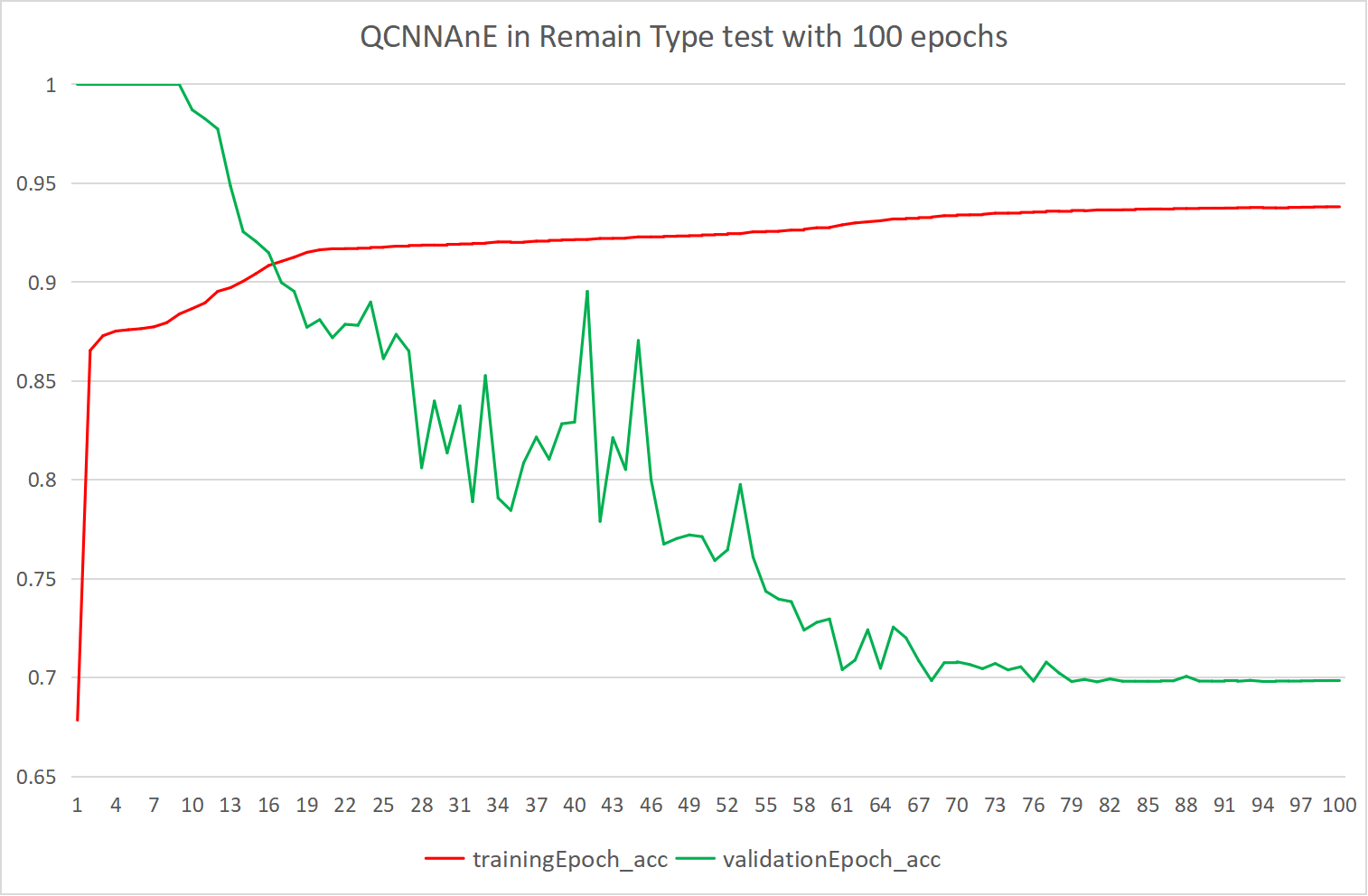}}
\caption{QCNNAnE in Remain Type test with 100 epochs}
\end{figure}

The reason we need to record the trade-off point of the train and test set accuracy is that all data in the test set is malicious and belong to only one type. It means the test set has a similar data pattern, thus, the whole data in the test set has high possibility to be recognized as malicious or benign. As a consequence, there is a probability of encountering extreme scenarios where the validation performance during the initial epoch may reach either 100\% or 0\%. However, as the model learned from the training set, this led to interference in the predictions for the test set. This interference became increasingly evident as the model delved deeper into the data structure of the training set (over-fitting). This kind of situation is more likely to occur when there is low data correlation between the test set and the training set. Thus, to record the trade-off accuracy between train and test set is worth to be analyzed.

Based on Figure 14, we found that as the number of epochs increases from 30 to 50, the accuracy of test set is still decreasing, and the accuracy of train set is gradually increasing. To find out whether the accuracy of train set and test set will keep steady in certain accuracy range finally, we continue to increase the number of epochs. We further extended the number of epochs from 50 to 100 so as to check whether the test accuracy will stop decreasing after certain epochs. we found that after 100 epochs, the performance of the model was unchanged already. Specifically, after 80 epochs, the test accuracy keeps steady at around 70\% and train accuracy keeps at around 93.5\%. Validation loss and training loss of this experiment is also plotted in Figure 15. After 20 epochs, the model has started to over-fit (the validation loss started to increase). If we select to early stop at trade-off point before 20 epochs, we can keep both training accuracy and test accuracy above 90\%.

\begin{figure}[h]
\centerline{\includegraphics[width=21pc]{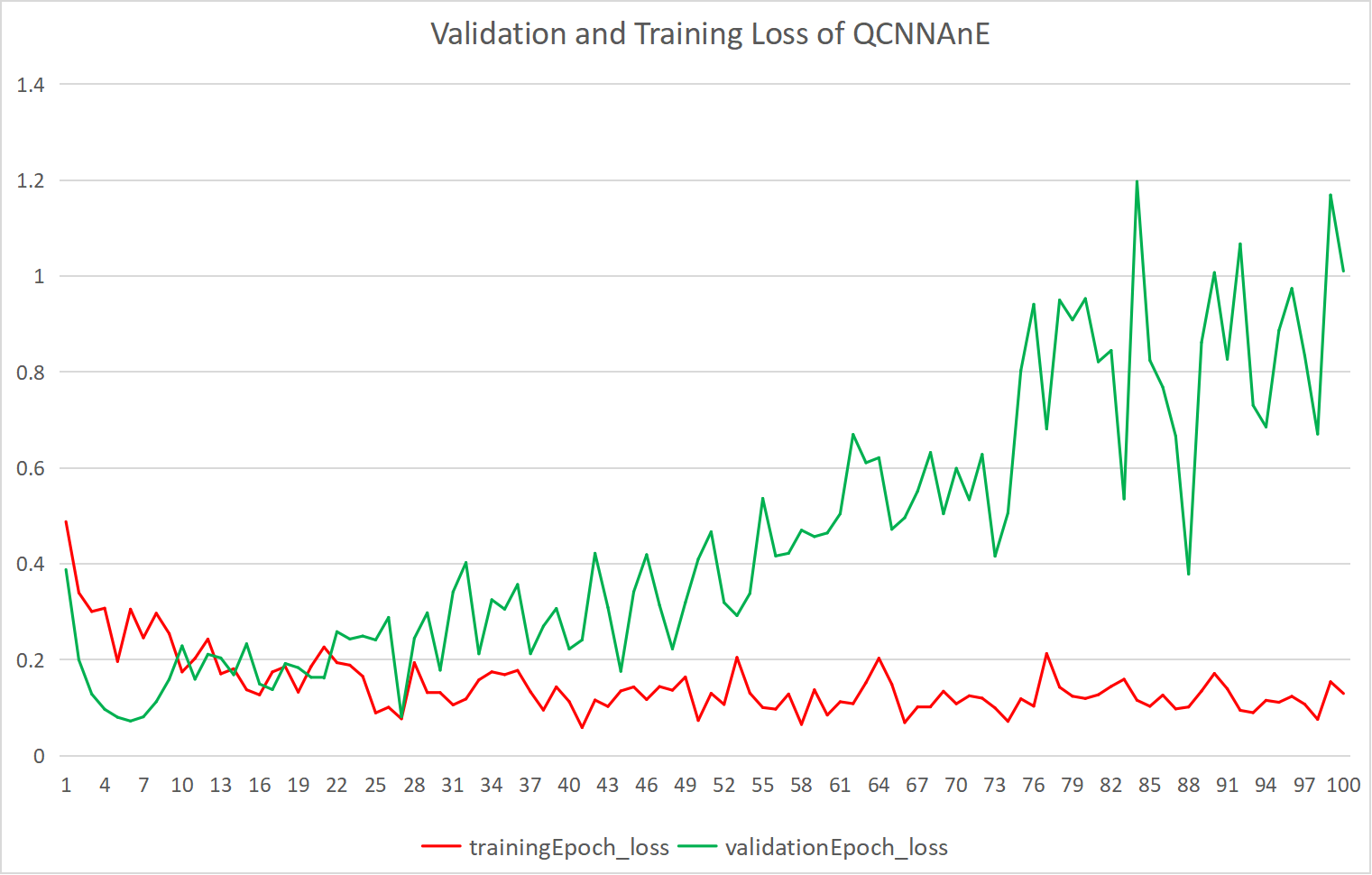}}
\caption{Validation and training loss of QCNNAnE in Remain Type test}
\end{figure}

Each model of each attack type is tested in 5 times, the average experiment results are recorded in Table IX and detailed results can be found in Appendix Experiment 3.

The experiment results presented in Table IX reveal substantial variations in the performance of different models across various test scenarios. Our evaluation of model performance encompasses two key aspects: the detection efficacy of known attack types and the ability to discern novel attack types. In Table IX, the Unseen Attack Type refers to the attack type out of the training set. Existing Attacks Type refers to attack types that appear in the model training set. Regarding the performance of existing attack types, as indicated by Table IX and Table X, virtually all models demonstrated a detection accuracy exceeding 83\%.

\begin{table*}[]
\centering
\caption{The Accuracy performance of different attack types in the final epoch in 5G NIDD dataset.}
\begin{tabular}{|c|cc|cc|cc|cc|}
\hline
               & \multicolumn{2}{c|}{\begin{tabular}[c]{@{}c@{}}Unseen Attack Type: \\ UDP Flood\end{tabular}}                                                   & \multicolumn{2}{c|}{\begin{tabular}[c]{@{}c@{}}Unseen Attack Type:\\ Remain Type\end{tabular}}                                                  & \multicolumn{2}{c|}{\begin{tabular}[c]{@{}c@{}}Unseen Attack Type:\\ HTTPFlood\end{tabular}}                                                    & \multicolumn{2}{c|}{\begin{tabular}[c]{@{}c@{}}Unseen Attack Type:\\ Slowrate DoS\end{tabular}}                                                 \\ \hline
               & \multicolumn{1}{c|}{\begin{tabular}[c]{@{}c@{}}Unseen \\ Attack\end{tabular}} & \begin{tabular}[c]{@{}c@{}}Existing \\ Attack\end{tabular} & \multicolumn{1}{c|}{\begin{tabular}[c]{@{}c@{}}Unseen \\ Attack\end{tabular}} & \begin{tabular}[c]{@{}c@{}}Existing \\ Attack\end{tabular} & \multicolumn{1}{c|}{\begin{tabular}[c]{@{}c@{}}Unseen \\ Attack\end{tabular}} & \begin{tabular}[c]{@{}c@{}}Existing \\ Attack\end{tabular} & \multicolumn{1}{c|}{\begin{tabular}[c]{@{}c@{}}Unseen \\ Attack\end{tabular}} & \begin{tabular}[c]{@{}c@{}}Existing \\ Attack\end{tabular} \\ \hline
CNN            & \multicolumn{1}{c|}{9.52\%}                                                & 99.70\%                                                    & \multicolumn{1}{c|}{83.72\%}                                               & 92.07\%                                                    & \multicolumn{1}{c|}{55.63\%}                                               & 92.08\%                                                    & \multicolumn{1}{c|}{79.85\%}                                               & 92.76\%                                                    \\ \hline
QCNNAnE        & \multicolumn{1}{c|}{15.19\%}                                               & 99.68\%                                                    & \multicolumn{1}{c|}{76.28\%}                                               & 92.41\%                                                    & \multicolumn{1}{c|}{59.33\%}                                               & 91.15\%                                                    & \multicolumn{1}{c|}{82.90\%}                                               & 92.47\%                                                    \\ \hline
QCNNAmE        & \multicolumn{1}{c|}{36.38\%}                                               & 95.36\%                                                    & \multicolumn{1}{c|}{84.10\%}                                               & 91.00\%                                                    & \multicolumn{1}{c|}{45.14\%}                                               & 91.61\%                                                    & \multicolumn{1}{c|}{82.05\%}                                               & 91.30\%                                                    \\ \hline
Quan-ConvCNN   & \multicolumn{1}{c|}{13.62\%}                                               & 99.61\%                                                    & \multicolumn{1}{c|}{87.84\%}                                               & 92.29\%                                                    & \multicolumn{1}{c|}{51.64\%}                                               & 91.66\%                                                    & \multicolumn{1}{c|}{75.00\%}                                               & 92.21\%                                                    \\ \hline
QuanvolutionNN & \multicolumn{1}{c|}{24.52\%}                                               & 99.20\%                                                    & \multicolumn{1}{c|}{87.27\%}                                               & 90.76\%                                                    & \multicolumn{1}{c|}{58.88\%}                                               & 90.14\%                                                    & \multicolumn{1}{c|}{88.30\%}                                               & 90.12\%                                                    \\ \hline
\end{tabular}
\end{table*}

\subsubsection{\bf \textbf{Experimental performance in the final 30 epoch}}

For the Remain Type test, except QCNNAnE model, other QCNN models outperform the classical CNN model. Quan-ConvCNN models achieved 87.84\% accuracy which is 4.12\% higher than the accuracy of classicial CNN.

In the context of the SlowrateDoS test, QCNNAnE, QCNNAmE, and QuanvolutionNN outperformed the classical CNN, with QuanvolutionNN achieving an 8\% higher performance. Although QuanvolutionNN attained the highest performance in the SlowrateDoS Test, its performance on existing attacks in Table IX and trade-off performances in Table X are all lower than other QCNN models.

In the HTTPFlood test, QCNNAnE and QuanvolutionNN demonstrated superior performance when compared to the CNN model. QCNNAnE model achieved a 3.3\% higher than the classical CNN model.

For the UDPFlood test, concerning the experimental performance in the final epoch (30th) in Table IX, all QCNN models demonstrated superior results compared to the classical CNN in the UDPFlood test. 

However, it is noteworthy that none of the models attained a detection rate exceeding 50\% in the UDPFlood test. Conversely, in both the Remain Types test and SlowrateDoS test, certain models exhibited performance levels above 85\%, such as Quan-ConvCNN and QuanvolutionNN models. This variability in performance can be attributed to differences in the volume and data structure of traffic associated with distinct attack types.

\begin{table*}[]
\centering
\caption{The Accuracy performance of different attack types test at the trade-off point in 5G NIDD dataset.}
\begin{tabular}{|c|cc|cc|cc|cc|}
\hline
               & \multicolumn{2}{c|}{\begin{tabular}[c]{@{}c@{}}Unseen Attack Type: \\ UDP Flood\end{tabular}}                                                   & \multicolumn{2}{c|}{\begin{tabular}[c]{@{}c@{}}Unseen Attack Type:\\ Remain Type\end{tabular}}                                                  & \multicolumn{2}{c|}{\begin{tabular}[c]{@{}c@{}}Unseen Attack Type:\\ HTTPFlood\end{tabular}}                                                    & \multicolumn{2}{c|}{\begin{tabular}[c]{@{}c@{}}Unseen Attack Type:\\ Slowrate DoS\end{tabular}}                                                 \\ \hline
               & \multicolumn{1}{c|}{\begin{tabular}[c]{@{}c@{}}Unseen \\ Attack\end{tabular}} & \begin{tabular}[c]{@{}c@{}}Existing \\ Attack\end{tabular} & \multicolumn{1}{c|}{\begin{tabular}[c]{@{}c@{}}Unseen \\ Attack\end{tabular}} & \begin{tabular}[c]{@{}c@{}}Existing \\ Attack\end{tabular} & \multicolumn{1}{c|}{\begin{tabular}[c]{@{}c@{}}Unseen \\ Attack\end{tabular}} & \begin{tabular}[c]{@{}c@{}}Existing \\ Attack\end{tabular} & \multicolumn{1}{c|}{\begin{tabular}[c]{@{}c@{}}Unseen \\ Attack\end{tabular}} & \begin{tabular}[c]{@{}c@{}}Existing \\ Attack\end{tabular} \\ \hline
CNN            & \multicolumn{1}{c|}{32.82\%}                                               & 92.05\%                                                    & \multicolumn{1}{c|}{91.44\%}                                               & 91.46\%                                                    & \multicolumn{1}{c|}{65.78\%}                                               & 87.63\%                                                    & \multicolumn{1}{c|}{91.82\%}                                               & 91.78\%                                                    \\ \hline
QCNNAnE        & \multicolumn{1}{c|}{18.47\%}                                               & 98.55\%                                                    & \multicolumn{1}{c|}{91.75\%}                                               & 91.59\%                                                    & \multicolumn{1}{c|}{67.56\%}                                               & 87.87\%                                                    & \multicolumn{1}{c|}{92.45\%}                                               & 92.32\%                                                    \\ \hline
QCNNAmE        & \multicolumn{1}{c|}{48.55\%}                                               & 96.40\%                                                    & \multicolumn{1}{c|}{93.64\%}                                               & 90.73\%                                                    & \multicolumn{1}{c|}{58.90\%}                                               & 83.51\%                                                    & \multicolumn{1}{c|}{90.86\%}                                               & 90.17\%                                                    \\ \hline
Quan-ConvCNN   & \multicolumn{1}{c|}{30.12\%}                                               & 96.08\%                                                    & \multicolumn{1}{c|}{92.31\%}                                               & 91.82\%                                                    & \multicolumn{1}{c|}{66.44\%}                                               & 87.19\%                                                    & \multicolumn{1}{c|}{91.08\%}                                               & 90.99\%                                                    \\ \hline
QuanvolutionNN & \multicolumn{1}{c|}{41.83\%}                                               & 85.93\%                                                    & \multicolumn{1}{c|}{91.85\%}                                               & 90.58\%                                                    & \multicolumn{1}{c|}{67.78\%}                                               & 85.77\%                                                    & \multicolumn{1}{c|}{90.15\%}                                               & 90.05\%                                                    \\ \hline
\end{tabular}
\end{table*}

\subsubsection{\bf \textbf{Experimental performance at the trade-off point}}

Regarding the results obtained at the trade-off point in Table X, QCNNAmE and QuanvolutionNN exhibit the top two highest accuracy performances in the UDPFlood test.

For the Remain Type Test, all QCNN models outperform the performance of the classical CNN model. QCNNAmE achieves the highest performance result which is 2.2\% higher than the classical CNN. 

While QuanvolutionNN achieved the highest performance in the HTTPFlood test, the margin is only 0.22\% higher than QCNNAnE. Notably, its performance on existing attacks is 2.1\% lower than QCNNAnE and 1.86\% lower than classical CNN. Consequently, QCNNAnE emerges as the most balanced and optimal trade-off model in the context of the HTTPFlood test. 

In the SlowrateDoS test, QCNNAnE achieved 92.45\% accuracy which is the highest performance among other models. 

Based on all the above comprehensive experimental results in experiment 3, the QCNN model demonstrates superior detection capabilities for unknown attack types compared to the classical CNN.

\subsection{\bf \textbf{Final Summary:}}

The results from Experiment 1 indicate that most Hybrid QCNN models outperform the Classical CNN in traditional offline ML. The experiment 1 also shows QCNN model with amplitude embedding method often exhibits significant oscillations during training, especially when embedding a large amount of information into a small number of qubits, leading to instability. In contrast, the classical CNN demonstrates a relatively more stable training process. A careful management of qubits allocation and embedding methods to mitigate training instabilities is required. For experiment 2, experiment results indicate that QCNN models have better robustness in facing unseen incoming traffic data. However, the experiment also shows that most QCNN models perform worse than classical CNN on small size datasets. The impact of insufficient training data is more pronounced on QCNN models compared to the
classical CNN. The extensive experimental results in Experiment 3 indicate that the QCNN model has better generalization and detection capabilities in the face of emerging attack types compared to the classical CNN. Although the optimal QCNN models varied across different tests, this demonstrates the potential of QCNN in this domain and points to the direction of further research into optimizing QCNN models. 

Overall, Hybrid QCNN models outperform the classical CNN in terms of accuracy and convergence speed during model training, particularly in handling unseen traffic and new emerging attack types.

\section{\bf Conclusion}

The study investigates the potential of hybrid QML in the domain of malicious traffic detection. To this end, we first provide a literature review of the current developments in QML in the field of malicious traffic detection. Subsequently, we assess whether the performance of hybrid QML in malicious traffic detection outperforms that of classical ML. Various architectures of QCNN models are designed, each employing distinct quantum embedding methods, quantum computing algorithms, and positions of quantum layers within the hybrid model. Following this, experimental comparisons are conducted to validate and contrast the performance of QCNN models against classical CNN models. Finally, an evaluation is conducted on the generalization and robustness of QCNN models on unknown traffic data from new data sources and new attack types. The experimental results demonstrate that QCNN models exhibit better performance in terms of detection accuracy, generalization, and robustness compared to classical CNN models. However, it is acknowledged that QML still requires further in-depth research in various aspects. For example, the initialization of parameters on qubits significantly impacts the convergence speed and final performance of the model. This will be our future research direction.

\section*{Declarations}
The author(s) received no financial support for the research, authorship, and/or publication of this article.

\onecolumn
\section*{Appendix}

\begin{table}[h!]
\centering
\caption{Experiment 1}
\begin{tabular}{|c|c|c|c|c|}
\hline
{  Detection Model} & {  1st}      & {  2nd}      & {  3rd}      & {  Average}  \\ \hline
{  CNN}             & {  99.62 \%} & {  99.52 \%} & {  99.33 \%} & {  99.49 \%} \\ \hline
{  QCNNAnE}         & {  99.69 \%} & {  99.60 \%} & {  99.57 \%} & {  99.62 \%} \\ \hline
{  QCNNAmE}         & {  99.73 \%} & {  99.69 \%} & {  99.52 \%} & {  99.65 \%} \\ \hline
{  QCNNMlayer}      & {  99.54 \%} & {  97.59 \%} & {  99.54 \%} & {  98.89 \%} \\ \hline
{  Quan-ConvCNN}    & {  99.73 \%} & {  99.76 \%} & {  99.66 \%} & {  99.71 \%} \\ \hline
{  QuanvolutionNN}  & {  98.82 \%} & {  98.67 \%} & {  98.53\%}  & {  98.67 \%} \\ \hline
\end{tabular}
\end{table}

\begin{table}[h!]
\centering
\caption{Experiment 2.1}
\begin{tabular}{|c|c|c|c|c|}
\hline
{  Detection Model} & {  1st}      & {  2nd}      & {  3rd}      & {  Average}  \\ \hline
{  CNN}             & {  78.94 \%} & {  78.71 \%} & {  77.05 \%} & {  78.23 \%} \\ \hline
{  QCNNAnE}         & {  77.66 \%} & {  78.66 \%} & {  78.59 \%} & {  78.30 \%} \\ \hline
{  QCNNAmE}         & {  78.49 \%} & {  78.96 \%} & {  78.98 \%} & {  78.81 \%} \\ \hline
{  QCNNMlayer}      & {  78.37 \%} & {  78.67 \%} & {  78.57 \%} & {  78.54 \%} \\ \hline
{  Quan-ConvCNN}    & {  79.04 \%} & {  78.91 \%} & {  78.86 \%} & {  78.94 \%} \\ \hline
{  QuanvolutionNN}  & {  77.96 \%} & {  79.25 \%} & {  78.76 \%} & {  78.66 \%} \\ \hline
\end{tabular}
\end{table}

\begin{table}[h!]
\centering
\caption{Experiment 2.2}
\begin{tabular}{|c|c|c|c|c|}
\hline
{  Detection Model} & {  1st}      & {  2nd}      & {  3rd}      & {  Average}  \\ \hline
{  CNN}             & {  75.23 \%} & {  69.91 \%} & {  77.71 \%} & {  74.28 \%} \\ \hline
{  QCNNAnE}         & {  58.47 \%} & {  58.53 \%} & {  61.08 \%} & {  59.36 \%} \\ \hline
{  QCNNAmE}         & {  84.06 \%} & {  81.66\%}  & {  78.98 \%} & {  81.57 \%} \\ \hline
{  QCNNMlayer}      & {  74.82 \%} & {  63.07 \%} & {  71.21 \%} & {  69.70 \%} \\ \hline
{  Quan-ConvCNN}    & {  69.60 \%} & {  58.28 \%} & {  64.07 \%} & {  64.07 \%} \\ \hline
{  QuanvolutionNN}  & {  57.44 \%} & {  57.07 \%} & {  57.02\%}  & {  57.18 \%} \\ \hline
\end{tabular}
\end{table}

\begin{table*}[h!]
\centering
\caption{Experiment 3: Remain Types (30 epochs)}
\begin{tabular}{|c|cc|cc|cc|cc|cc|}
\hline
{  }        & \multicolumn{2}{c|}{{  CNN}}                                      & \multicolumn{2}{c|}{{  QCNNAnE}}                                  & \multicolumn{2}{c|}{{  QCNNAmE}}                                  & \multicolumn{2}{c|}{{  Quan-ConvCNN}}                             & \multicolumn{2}{c|}{{  QuanvolutionNN}}                           \\ \hline
{  }        & \multicolumn{1}{c|}{{  Test}}    & {  Train}   & \multicolumn{1}{c|}{{  Test}}    & {  Train}   & \multicolumn{1}{c|}{{  Test}}    & {  Train}   & \multicolumn{1}{c|}{{  Test}}    & {  Train}   & \multicolumn{1}{c|}{{  Test}}    & {  Train}   \\ \hline
{  1}       & \multicolumn{1}{c|}{{  88.24\%}} & {  91.94\%} & \multicolumn{1}{c|}{{  73.41\%}} & {  92.14\%} & \multicolumn{1}{c|}{{  78.89\%}} & {  91.99\%} & \multicolumn{1}{c|}{{  93.78\%}} & {  92.28\%} & \multicolumn{1}{c|}{{  91.81\%}} & {  90.41\%} \\ \hline
{  2}       & \multicolumn{1}{c|}{{  75.36\%}} & {  92.07\%} & \multicolumn{1}{c|}{{  81.36\%}} & {  91.90\%} & \multicolumn{1}{c|}{{  82.67\%}} & {  91.06\%} & \multicolumn{1}{c|}{{  89.49\%}} & {  92.08\%} & \multicolumn{1}{c|}{{  92.69\%}} & {  90.61\%} \\ \hline
{  3}       & \multicolumn{1}{c|}{{  77.20\%}} & {  92.83\%} & \multicolumn{1}{c|}{{  68.32\%}} & {  93.73\%} & \multicolumn{1}{c|}{{  66.53\%}} & {  91.88\%} & \multicolumn{1}{c|}{{  87.32\%}} & {  92.39\%} & \multicolumn{1}{c|}{{  74.47\%}} & {  90.72\%} \\ \hline
{  4}       & \multicolumn{1}{c|}{{  87.51\%}} & {  92.61\%} & \multicolumn{1}{c|}{{  77.34\%}} & {  92.21\%} & \multicolumn{1}{c|}{{  98.16\%}} & {  89.51\%} & \multicolumn{1}{c|}{{  93.53\%}} & {  92.34\%} & \multicolumn{1}{c|}{{  84.50\%}} & {  91.01\%} \\ \hline
{  5}       & \multicolumn{1}{c|}{{  90.30\%}} & {  90.92\%} & \multicolumn{1}{c|}{{  80.95\%}} & {  92.09\%} & \multicolumn{1}{c|}{{  94.24\%}} & {  90.58\%} & \multicolumn{1}{c|}{{  75.06\%}} & {  92.37\%} & \multicolumn{1}{c|}{{  92.90\%}} & {  91.04\%} \\ \hline
{  Average} & \multicolumn{1}{c|}{{  83.72\%}} & {  92.07\%} & \multicolumn{1}{c|}{{  76.28\%}} & {  92.41\%} & \multicolumn{1}{c|}{{  84.10\%}} & {  91.00\%} & \multicolumn{1}{c|}{{  87.84\%}} & {  92.29\%} & \multicolumn{1}{c|}{{  87.27\%}} & {  90.76\%} \\ \hline
{  Median}  & \multicolumn{1}{c|}{{  87.51\%}} & {  92.61\%} & \multicolumn{1}{c|}{{  77.34\%}} & {  92.21\%} & \multicolumn{1}{c|}{{  82.67\%}} & {  91.06\%} & \multicolumn{1}{c|}{{  89.49\%}} & {  92.08\%} & \multicolumn{1}{c|}{{  91.81\%}} & {  90.41\%} \\ \hline
\end{tabular}
\end{table*}

\begin{table*}[h!]
\centering
\caption{Experiment 3: Remain Types (Trade-off)}
\begin{tabular}{|c|cc|cc|cc|cc|cc|}
\hline
{  }        & \multicolumn{2}{c|}{{  CNN}}                                      & \multicolumn{2}{c|}{{  QCNNAnE}}                                  & \multicolumn{2}{c|}{{  QCNNAmE}}                                  & \multicolumn{2}{c|}{{  Quan-ConvCNN}}                             & \multicolumn{2}{c|}{{  QuanvolutionNN}}                           \\ \hline
{  }        & \multicolumn{1}{c|}{{  Test}}    & {  Train}   & \multicolumn{1}{c|}{{  Test}}    & {  Train}   & \multicolumn{1}{c|}{{  Test}}    & {  Train}   & \multicolumn{1}{c|}{{  Test}}    & {  Train}   & \multicolumn{1}{c|}{{  Test}}    & {  Train}   \\ \hline
{  1}       & \multicolumn{1}{c|}{{  91.16\%}} & {  91.35\%} & \multicolumn{1}{c|}{{  91.56\%}} & {  91.58\%} & \multicolumn{1}{c|}{{  91.97\%}} & {  90.75\%} & \multicolumn{1}{c|}{{  93.78\%}} & {  92.28\%} & \multicolumn{1}{c|}{{  91.81\%}} & {  90.41\%} \\ \hline
{  2}       & \multicolumn{1}{c|}{{  91.14\%}} & {  91.83\%} & \multicolumn{1}{c|}{{  91.48\%}} & {  90.83\%} & \multicolumn{1}{c|}{{  95.35\%}} & {  90.66\%} & \multicolumn{1}{c|}{{  91.94\%}} & {  92.01\%} & \multicolumn{1}{c|}{{  92.69\%}} & {  90.61\%} \\ \hline
{  3}       & \multicolumn{1}{c|}{{  91.95\%}} & {  91.41\%} & \multicolumn{1}{c|}{{  92.04\%}} & {  91.75\%} & \multicolumn{1}{c|}{{  91.86\%}} & {  91.72\%} & \multicolumn{1}{c|}{{  91.68\%}} & {  92.00\%} & \multicolumn{1}{c|}{{  91.46\%}} & {  90.08\%} \\ \hline
{  4}       & \multicolumn{1}{c|}{{  91.65\%}} & {  91.77\%} & \multicolumn{1}{c|}{{  92.17\%}} & {  92.01\%} & \multicolumn{1}{c|}{{  98.16\%}} & {  89.51\%} & \multicolumn{1}{c|}{{  93.53\%}} & {  92.34\%} & \multicolumn{1}{c|}{{  90.41\%}} & {  90.76\%} \\ \hline
{  5}       & \multicolumn{1}{c|}{{  91.32\%}} & {  90.95\%} & \multicolumn{1}{c|}{{  91.50\%}} & {  91.77\%} & \multicolumn{1}{c|}{{  90.88\%}} & {  91.00\%} & \multicolumn{1}{c|}{{  90.60\%}} & {  90.46\%} & \multicolumn{1}{c|}{{  92.90\%}} & {  91.04\%} \\ \hline
{  Average} & \multicolumn{1}{c|}{{  91.44\%}} & {  91.46\%} & \multicolumn{1}{c|}{{  91.75\%}} & {  91.59\%} & \multicolumn{1}{c|}{{  93.64\%}} & {  90.73\%} & \multicolumn{1}{c|}{{  92.31\%}} & {  91.82\%} & \multicolumn{1}{c|}{{  91.85\%}} & {  90.58\%} \\ \hline
{  Median}  & \multicolumn{1}{c|}{{  91.32\%}} & {  90.95\%} & \multicolumn{1}{c|}{{  91.56\%}} & {  91.58\%} & \multicolumn{1}{c|}{{  91.97\%}} & {  90.75\%} & \multicolumn{1}{c|}{{  91.94\%}} & {  92.01\%} & \multicolumn{1}{c|}{{  91.81\%}} & {  90.41\%} \\ \hline
\end{tabular}
\end{table*}

\begin{table*}[h!]
\centering
\caption{Experiment 3: UDP Flood (30 epochs)}
\begin{tabular}{|c|cc|cc|cc|cc|cc|}
\hline
{  }        & \multicolumn{2}{c|}{{  CNN}}                                      & \multicolumn{2}{c|}{{  QCNNAnE}}                                  & \multicolumn{2}{c|}{{  QCNNAmE}}                                  & \multicolumn{2}{c|}{{  Quan-ConvCNN}}                             & \multicolumn{2}{c|}{{  QuanvolutionNN}}                           \\ \hline
{  }        & \multicolumn{1}{c|}{{  Test}}    & {  Train}   & \multicolumn{1}{c|}{{  Test}}    & {  Train}   & \multicolumn{1}{c|}{{  Test}}    & {  Train}   & \multicolumn{1}{c|}{{  Test}}    & {  Train}   & \multicolumn{1}{c|}{{  Test}}    & {  Train}   \\ \hline
{  1}       & \multicolumn{1}{c|}{{  11.11\%}} & {  99.65\%} & \multicolumn{1}{c|}{{  16.16\%}} & {  99.64\%} & \multicolumn{1}{c|}{{  6.46\%}}  & {  97.61\%} & \multicolumn{1}{c|}{{  12.01\%}} & {  99.51\%} & \multicolumn{1}{c|}{{  19.52\%}} & {  99.59\%} \\ \hline
{  2}       & \multicolumn{1}{c|}{{  7.85\%}}  & {  99.88\%} & \multicolumn{1}{c|}{{  14.56\%}} & {  99.72\%} & \multicolumn{1}{c|}{{  34.91\%}} & {  99.51\%} & \multicolumn{1}{c|}{{  10.72\%}} & {  99.65\%} & \multicolumn{1}{c|}{{  9.69\%}}  & {  98.98\%} \\ \hline
{  3}       & \multicolumn{1}{c|}{{  8.42\%}}  & {  99.51\%} & \multicolumn{1}{c|}{{  11.69\%}} & {  99.80\%} & \multicolumn{1}{c|}{{  69.56\%}} & {  80.58\%} & \multicolumn{1}{c|}{{  19.93\%}} & {  99.66\%} & \multicolumn{1}{c|}{{  12.50\%}} & {  98.92\%} \\ \hline
{  4}       & \multicolumn{1}{c|}{{  11.94\%}} & {  99.58\%} & \multicolumn{1}{c|}{{  16.90\%}} & {  99.47\%} & \multicolumn{1}{c|}{{  62.90\%}} & {  99.40\%} & \multicolumn{1}{c|}{{  7.90\%}}  & {  99.66\%} & \multicolumn{1}{c|}{{  36.28\%}} & {  99.52\%} \\ \hline
{  5}       & \multicolumn{1}{c|}{{  8.26\%}}  & {  99.86\%} & \multicolumn{1}{c|}{{  16.66\%}} & {  99.76\%} & \multicolumn{1}{c|}{{  8.05\%}}  & {  99.71\%} & \multicolumn{1}{c|}{{  17.52\%}} & {  99.58\%} & \multicolumn{1}{c|}{{  44.60\%}} & {  98.98\%} \\ \hline
{  Average} & \multicolumn{1}{c|}{{  9.52\%}}  & {  99.70\%} & \multicolumn{1}{c|}{{  15.19\%}} & {  99.68\%} & \multicolumn{1}{c|}{{  36.38\%}} & {  95.36\%} & \multicolumn{1}{c|}{{  13.62\%}} & {  99.61\%} & \multicolumn{1}{c|}{{  24.52\%}} & {  99.20\%} \\ \hline
{  Median}  & \multicolumn{1}{c|}{{  8.42\%}}  & {  99.51\%} & \multicolumn{1}{c|}{{  16.16\%}} & {  99.64\%} & \multicolumn{1}{c|}{{  34.91\%}} & {  99.51\%} & \multicolumn{1}{c|}{{  12.01\%}} & {  99.51\%} & \multicolumn{1}{c|}{{  36.28\%}} & {  99.52\%} \\ \hline
\end{tabular}
\end{table*}

\begin{table*}[h!]
\centering
\caption{Experiment 3: UDP Flood (Trade-Off)}
\begin{tabular}{|c|cc|cc|cc|cc|cc|}
\hline
{  }        & \multicolumn{2}{c|}{{  CNN}}                                      & \multicolumn{2}{c|}{{  QCNNAnE}}                                  & \multicolumn{2}{c|}{{  QCNNAmE}}                                  & \multicolumn{2}{c|}{{  Quan-ConvCNN}}                             & \multicolumn{2}{c|}{{  QuanvolutionNN}}                           \\ \hline
{  }        & \multicolumn{1}{c|}{{  Test}}    & {  Train}   & \multicolumn{1}{c|}{{  Test}}    & {  Train}   & \multicolumn{1}{c|}{{  Test}}    & {  Train}   & \multicolumn{1}{c|}{{  Test}}    & {  Train}   & \multicolumn{1}{c|}{{  Test}}    & {  Train}   \\ \hline
{  1}       & \multicolumn{1}{c|}{{  35.05\%}} & {  93.61\%} & \multicolumn{1}{c|}{{  19.88\%}} & {  97.30\%} & \multicolumn{1}{c|}{{  9.35\%}}  & {  97.55\%} & \multicolumn{1}{c|}{{  53.04\%}} & {  92.74\%} & \multicolumn{1}{c|}{{  28.00\%}} & {  80.90\%} \\ \hline
{  2}       & \multicolumn{1}{c|}{{  21.86\%}} & {  85.04\%} & \multicolumn{1}{c|}{{  14.56\%}} & {  99.72\%} & \multicolumn{1}{c|}{{  59.39\%}} & {  96.90\%} & \multicolumn{1}{c|}{{  31.85\%}} & {  94.41\%} & \multicolumn{1}{c|}{{  35.12\%}} & {  77.09\%} \\ \hline
{  3}       & \multicolumn{1}{c|}{{  16.49\%}} & {  94.38\%} & \multicolumn{1}{c|}{{  13.23\%}} & {  96.77\%} & \multicolumn{1}{c|}{{  87.49\%}} & {  91.86\%} & \multicolumn{1}{c|}{{  19.93\%}} & {  99.66\%} & \multicolumn{1}{c|}{{  45.63\%}} & {  84.82\%} \\ \hline
{  4}       & \multicolumn{1}{c|}{{  43.30\%}} & {  92.79\%} & \multicolumn{1}{c|}{{  26.40\%}} & {  99.29\%} & \multicolumn{1}{c|}{{  76.62\%}} & {  96.00\%} & \multicolumn{1}{c|}{{  27.26\%}} & {  94.02\%} & \multicolumn{1}{c|}{{  55.78\%}} & {  93.36\%} \\ \hline
{  5}       & \multicolumn{1}{c|}{{  47.42\%}} & {  94.45\%} & \multicolumn{1}{c|}{{  18.27\%}} & {  99.69\%} & \multicolumn{1}{c|}{{  9.90\%}}  & {  99.68\%} & \multicolumn{1}{c|}{{  18.50\%}} & {  99.58\%} & \multicolumn{1}{c|}{{  44.61\%}} & {  93.48\%} \\ \hline
{  Average} & \multicolumn{1}{c|}{{  32.82\%}} & {  92.05\%} & \multicolumn{1}{c|}{{  18.47\%}} & {  98.55\%} & \multicolumn{1}{c|}{{  48.55\%}} & {  96.40\%} & \multicolumn{1}{c|}{{  30.12\%}} & {  96.08\%} & \multicolumn{1}{c|}{{  41.83\%}} & {  85.93\%} \\ \hline
{  Median}  & \multicolumn{1}{c|}{{  35.05\%}} & {  93.61\%} & \multicolumn{1}{c|}{{  18.27\%}} & {  99.69\%} & \multicolumn{1}{c|}{{  59.39\%}} & {  96.90\%} & \multicolumn{1}{c|}{{  27.26\%}} & {  94.02\%} & \multicolumn{1}{c|}{{  44.61\%}} & {  93.48\%} \\ \hline
\end{tabular}
\end{table*}

\begin{table*}[h!]
\centering
\caption{Experiment 3: SlowrateDoS (30 epochs)}
\begin{tabular}{|c|cc|cc|cc|cc|cc|}
\hline
{  }        & \multicolumn{2}{c|}{{  CNN}}                                      & \multicolumn{2}{c|}{{  QCNNAnE}}                                  & \multicolumn{2}{c|}{{  QCNNAmE}}                                  & \multicolumn{2}{c|}{{  Quan-ConvCNN}}                             & \multicolumn{2}{c|}{{  QuanvolutionNN}}                           \\ \hline
{  }        & \multicolumn{1}{c|}{{  Test}}    & {  Train}   & \multicolumn{1}{c|}{{  Test}}    & {  Train}   & \multicolumn{1}{c|}{{  Test}}    & {  Train}   & \multicolumn{1}{c|}{{  Test}}    & {  Train}   & \multicolumn{1}{c|}{{  Test}}    & {  Train}   \\ \hline
{  1}       & \multicolumn{1}{c|}{{  69.70\%}} & {  92.23\%} & \multicolumn{1}{c|}{{  90.48\%}} & {  93.29\%} & \multicolumn{1}{c|}{{  80.57\%}} & {  90.70\%} & \multicolumn{1}{c|}{{  85.96\%}} & {  92.87\%} & \multicolumn{1}{c|}{{  80.93\%}} & {  91.02\%} \\ \hline
{  2}       & \multicolumn{1}{c|}{{  84.56\%}} & {  92.99\%} & \multicolumn{1}{c|}{{  94.90\%}} & {  93.84\%} & \multicolumn{1}{c|}{{  85.24\%}} & {  92.10\%} & \multicolumn{1}{c|}{{  51.30\%}} & {  92.15\%} & \multicolumn{1}{c|}{{  88.99\%}} & {  89.81\%} \\ \hline
{  3}       & \multicolumn{1}{c|}{{  68.37\%}} & {  93.86\%} & \multicolumn{1}{c|}{{  72.42\%}} & {  90.97\%} & \multicolumn{1}{c|}{{  78.11\%}} & {  91.78\%} & \multicolumn{1}{c|}{{  85.11\%}} & {  91.54\%} & \multicolumn{1}{c|}{{  90.89\%}} & {  90.00\%} \\ \hline
{  4}       & \multicolumn{1}{c|}{{  91.83\%}} & {  91.92\%} & \multicolumn{1}{c|}{{  62.78\%}} & {  90.96\%} & \multicolumn{1}{c|}{{  76.70\%}} & {  90.91\%} & \multicolumn{1}{c|}{{  68.76\%}} & {  92.21\%} & \multicolumn{1}{c|}{{  90.92\%}} & {  89.71\%} \\ \hline
{  5}       & \multicolumn{1}{c|}{{  84.78\%}} & {  92.82\%} & \multicolumn{1}{c|}{{  93.92\%}} & {  93.31\%} & \multicolumn{1}{c|}{{  89.63\%}} & {  91.03\%} & \multicolumn{1}{c|}{{  83.84\%}} & {  92.29\%} & \multicolumn{1}{c|}{{  89.79\%}} & {  90.04\%} \\ \hline
{  Average} & \multicolumn{1}{c|}{{  79.85\%}} & {  92.76\%} & \multicolumn{1}{c|}{{  82.90\%}} & {  92.47\%} & \multicolumn{1}{c|}{{  82.05\%}} & {  91.30\%} & \multicolumn{1}{c|}{{  75.00\%}} & {  92.21\%} & \multicolumn{1}{c|}{{  88.30\%}} & {  90.12\%} \\ \hline
{  Median}  & \multicolumn{1}{c|}{{  84.56\%}} & {  92.99\%} & \multicolumn{1}{c|}{{  90.48\%}} & {  93.29\%} & \multicolumn{1}{c|}{{  80.57\%}} & {  90.70\%} & \multicolumn{1}{c|}{{  83.84\%}} & {  92.29\%} & \multicolumn{1}{c|}{{  89.79\%}} & {  90.04\%} \\ \hline
\end{tabular}
\end{table*}

\begin{table*}[h!]
\centering
\caption{Experiment 3: SlowrateDoS (Trade-Off)}
\begin{tabular}{|c|cc|cc|cc|cc|cc|}
\hline
{  }        & \multicolumn{2}{c|}{{  CNN}}                                      & \multicolumn{2}{c|}{{  QCNNAnE}}                                  & \multicolumn{2}{c|}{{  QCNNAmE}}                                  & \multicolumn{2}{c|}{{  Quan-ConvCNN}}                             & \multicolumn{2}{c|}{{  QuanvolutionNN}}                           \\ \hline
{  }        & \multicolumn{1}{c|}{{  Test}}    & {  Train}   & \multicolumn{1}{c|}{{  Test}}    & {  Train}   & \multicolumn{1}{c|}{{  Test}}    & {  Train}   & \multicolumn{1}{c|}{{  Test}}    & {  Train}   & \multicolumn{1}{c|}{{  Test}}    & {  Train}   \\ \hline
{  1}       & \multicolumn{1}{c|}{{  92.06\%}} & {  91.39\%} & \multicolumn{1}{c|}{{  91.96\%}} & {  92.79\%} & \multicolumn{1}{c|}{{  89.35\%}} & {  90.68\%} & \multicolumn{1}{c|}{{  92.53\%}} & {  92.14\%} & \multicolumn{1}{c|}{{  90.18\%}} & {  90.69\%} \\ \hline
{  2}       & \multicolumn{1}{c|}{{  92.53\%}} & {  92.44\%} & \multicolumn{1}{c|}{{  93.70\%}} & {  93.75\%} & \multicolumn{1}{c|}{{  91.73\%}} & {  91.44\%} & \multicolumn{1}{c|}{{  90.27\%}} & {  90.15\%} & \multicolumn{1}{c|}{{  88.99\%}} & {  89.81\%} \\ \hline
{  3}       & \multicolumn{1}{c|}{{  91.23\%}} & {  91.64\%} & \multicolumn{1}{c|}{{  91.15\%}} & {  90.85\%} & \multicolumn{1}{c|}{{  91.23\%}} & {  88.49\%} & \multicolumn{1}{c|}{{  91.25\%}} & {  91.21\%} & \multicolumn{1}{c|}{{  90.89\%}} & {  90.00\%} \\ \hline
{  4}       & \multicolumn{1}{c|}{{  91.83\%}} & {  91.92\%} & \multicolumn{1}{c|}{{  91.52\%}} & {  90.91\%} & \multicolumn{1}{c|}{{  90.87\%}} & {  89.96\%} & \multicolumn{1}{c|}{{  89.21\%}} & {  89.65\%} & \multicolumn{1}{c|}{{  90.92\%}} & {  89.71\%} \\ \hline
{  5}       & \multicolumn{1}{c|}{{  91.45\%}} & {  91.47\%} & \multicolumn{1}{c|}{{  93.92\%}} & {  93.31\%} & \multicolumn{1}{c|}{{  91.11\%}} & {  90.26\%} & \multicolumn{1}{c|}{{  92.13\%}} & {  91.81\%} & \multicolumn{1}{c|}{{  89.79\%}} & {  90.04\%} \\ \hline
{  Average} & \multicolumn{1}{c|}{{  91.82\%}} & {  91.78\%} & \multicolumn{1}{c|}{{  92.45\%}} & {  92.32\%} & \multicolumn{1}{c|}{{  90.86\%}} & {  90.17\%} & \multicolumn{1}{c|}{{  91.08\%}} & {  90.99\%} & \multicolumn{1}{c|}{{  90.15\%}} & {  90.05\%} \\ \hline
{  Median}  & \multicolumn{1}{c|}{{  91.83\%}} & {  91.92\%} & \multicolumn{1}{c|}{{  91.96\%}} & {  92.79\%} & \multicolumn{1}{c|}{{  91.11\%}} & {  90.26\%} & \multicolumn{1}{c|}{{  91.25\%}} & {  91.21\%} & \multicolumn{1}{c|}{{  90.18\%}} & {  90.69\%} \\ \hline
\end{tabular}
\end{table*}

\begin{table*}[h!]
\centering
\caption{Experiment 3: HTTPFlood (30 epochs)}
\begin{tabular}{|c|cc|cc|cc|cc|cc|}
\hline
{  }        & \multicolumn{2}{c|}{{  CNN}}                                      & \multicolumn{2}{c|}{{  QCNNAnE}}                                  & \multicolumn{2}{c|}{{  QCNNAmE}}                                  & \multicolumn{2}{c|}{{  Quan-ConvCNN}}                             & \multicolumn{2}{c|}{{  QuanvolutionNN}}                           \\ \hline
{  }        & \multicolumn{1}{c|}{{  Test}}    & {  Train}   & \multicolumn{1}{c|}{{  Test}}    & {  Train}   & \multicolumn{1}{c|}{{  Test}}    & {  Train}   & \multicolumn{1}{c|}{{  Test}}    & {  Train}   & \multicolumn{1}{c|}{{  Test}}    & {  Train}   \\ \hline
{  1}       & \multicolumn{1}{c|}{{  60.83\%}} & {  92.26\%} & \multicolumn{1}{c|}{{  68.27\%}} & {  90.46\%} & \multicolumn{1}{c|}{{  49.13\%}} & {  93.27\%} & \multicolumn{1}{c|}{{  49.11\%}} & {  91.69\%} & \multicolumn{1}{c|}{{  51.16\%}} & {  91.18\%} \\ \hline
{  2}       & \multicolumn{1}{c|}{{  57.24\%}} & {  91.50\%} & \multicolumn{1}{c|}{{  45.56\%}} & {  90.46\%} & \multicolumn{1}{c|}{{  30.91\%}} & {  92.10\%} & \multicolumn{1}{c|}{{  54.48\%}} & {  91.63\%} & \multicolumn{1}{c|}{{  61.69\%}} & {  89.24\%} \\ \hline
{  3}       & \multicolumn{1}{c|}{{  61.11\%}} & {  92.80\%} & \multicolumn{1}{c|}{{  57.52\%}} & {  91.49\%} & \multicolumn{1}{c|}{{  32.97\%}} & {  91.64\%} & \multicolumn{1}{c|}{{  58.02\%}} & {  91.64\%} & \multicolumn{1}{c|}{{  58.16\%}} & {  89.93\%} \\ \hline
{  4}       & \multicolumn{1}{c|}{{  59.51\%}} & {  92.39\%} & \multicolumn{1}{c|}{{  63.56\%}} & {  91.50\%} & \multicolumn{1}{c|}{{  51.82\%}} & {  90.46\%} & \multicolumn{1}{c|}{{  51.45\%}} & {  91.37\%} & \multicolumn{1}{c|}{{  61.71\%}} & {  90.10\%} \\ \hline
{  5}       & \multicolumn{1}{c|}{{  39.46\%}} & {  91.43\%} & \multicolumn{1}{c|}{{  61.75\%}} & {  91.84\%} & \multicolumn{1}{c|}{{  60.85\%}} & {  90.58\%} & \multicolumn{1}{c|}{{  45.15\%}} & {  92.00\%} & \multicolumn{1}{c|}{{  61.68\%}} & {  90.23\%} \\ \hline
{  Average} & \multicolumn{1}{c|}{{  55.63\%}} & {  92.08\%} & \multicolumn{1}{c|}{{  59.33\%}} & {  91.15\%} & \multicolumn{1}{c|}{{  45.14\%}} & {  91.61\%} & \multicolumn{1}{c|}{{  51.64\%}} & {  91.66\%} & \multicolumn{1}{c|}{{  58.88\%}} & {  90.14\%} \\ \hline
{  Median}  & \multicolumn{1}{c|}{{  59.51\%}} & {  92.39\%} & \multicolumn{1}{c|}{{  61.75\%}} & {  91.84\%} & \multicolumn{1}{c|}{{  49.13\%}} & {  93.27\%} & \multicolumn{1}{c|}{{  51.45\%}} & {  91.37\%} & \multicolumn{1}{c|}{{  61.68\%}} & {  90.23\%} \\ \hline
\end{tabular}
\end{table*}

\begin{table*}[h!]
\centering
\caption{Experiment 3: HTTPFlood (Trade-Off)}
\begin{tabular}{|c|cc|cc|cc|cc|cc|}
\hline
{  }        & \multicolumn{2}{c|}{{  CNN}}                                      & \multicolumn{2}{c|}{{  QCNNAnE}}                                  & \multicolumn{2}{c|}{{  QCNNAmE}}                                  & \multicolumn{2}{c|}{{  Quan-ConvCNN}}                             & \multicolumn{2}{c|}{{  QuanvolutionNN}}                           \\ \hline
{  }        & \multicolumn{1}{c|}{{  Test}}    & {  Train}   & \multicolumn{1}{c|}{{  Test}}    & {  Train}   & \multicolumn{1}{c|}{{  Test}}    & {  Train}   & \multicolumn{1}{c|}{{  Test}}    & {  Train}   & \multicolumn{1}{c|}{{  Test}}    & {  Train}   \\ \hline
{  1}       & \multicolumn{1}{c|}{{  66.51\%}} & {  87.66\%} & \multicolumn{1}{c|}{{  70.72\%}} & {  90.29\%} & \multicolumn{1}{c|}{{  57.87\%}} & {  88.21\%} & \multicolumn{1}{c|}{{  66.81\%}} & {  87.36\%} & \multicolumn{1}{c|}{{  68.82\%}} & {  86.68\%} \\ \hline
{  2}       & \multicolumn{1}{c|}{{  62.23\%}} & {  87.96\%} & \multicolumn{1}{c|}{{  66.07\%}} & {  87.31\%} & \multicolumn{1}{c|}{{  51.43\%}} & {  64.71\%} & \multicolumn{1}{c|}{{  65.95\%}} & {  87.47\%} & \multicolumn{1}{c|}{{  75.14\%}} & {  81.43\%} \\ \hline
{  3}       & \multicolumn{1}{c|}{{  66.78\%}} & {  87.32\%} & \multicolumn{1}{c|}{{  66.85\%}} & {  87.23\%} & \multicolumn{1}{c|}{{  49.97\%}} & {  91.53\%} & \multicolumn{1}{c|}{{  67.18\%}} & {  86.67\%} & \multicolumn{1}{c|}{{  66.21\%}} & {  84.00\%} \\ \hline
{  4}       & \multicolumn{1}{c|}{{  66.40\%}} & {  87.66\%} & \multicolumn{1}{c|}{{  67.22\%}} & {  87.32\%} & \multicolumn{1}{c|}{{  67.01\%}} & {  85.95\%} & \multicolumn{1}{c|}{{  66.22\%}} & {  86.98\%} & \multicolumn{1}{c|}{{  61.71\%}} & {  90.10\%} \\ \hline
{  5}       & \multicolumn{1}{c|}{{  66.99\%}} & {  87.57\%} & \multicolumn{1}{c|}{{  66.96\%}} & {  87.21\%} & \multicolumn{1}{c|}{{  68.23\%}} & {  87.14\%} & \multicolumn{1}{c|}{{  66.03\%}} & {  87.49\%} & \multicolumn{1}{c|}{{  67.02\%}} & {  86.65\%} \\ \hline
{  Average} & \multicolumn{1}{c|}{{  65.78\%}} & {  87.63\%} & \multicolumn{1}{c|}{{  67.56\%}} & {  87.87\%} & \multicolumn{1}{c|}{{  58.90\%}} & {  83.51\%} & \multicolumn{1}{c|}{{  66.44\%}} & {  87.19\%} & \multicolumn{1}{c|}{{  67.78\%}} & {  85.77\%} \\ \hline
{  Median}  & \multicolumn{1}{c|}{{  66.51\%}} & {  87.66\%} & \multicolumn{1}{c|}{{  66.96\%}} & {  87.21\%} & \multicolumn{1}{c|}{{  57.87\%}} & {  88.21\%} & \multicolumn{1}{c|}{{  66.22\%}} & {  86.98\%} & \multicolumn{1}{c|}{{  67.10\%}} & {  76.60\%} \\ \hline
\end{tabular}
\end{table*}

\end{document}